\documentclass{aa}  

\usepackage{soul}
\usepackage{ulem}
\usepackage{cancel}
\usepackage{graphicx}   
\usepackage{amsmath}    
\usepackage{soul}
\usepackage{color}
\usepackage[dvipsnames]{xcolor}
\usepackage{natbib}
\usepackage{url}
\usepackage{placeins}
\usepackage{xfrac}
\usepackage{comment}
\usepackage[breaklinks=true]{hyperref} 
\usepackage{amssymb}
\usepackage{booktabs}
\usepackage{soul}
\usepackage{tabularx}
\usepackage{verbatim}
\usepackage{orcidlink}

\def\apj {ApJ}
\def\apjl {ApJL}
\def\apjs {ApJS}
\def\aj {AJ}
\def\aap {A\&A}
\def\mnras {MNRAS}
\def\araa {ARA\&A}

\def\R200 {R_{200}}

\def\Kpc {\rm Kpc}

\defcitealias{Palma:2025}{Paper I}

\DeclareRobustCommand{\VAN}[3]{#2}
\let\VANthebibliography\thebibliography
\def\thebibliography{\DeclareRobustCommand{\VAN}[3]{##3}\VANthebibliography}

\usepackage{txfonts}
\begin{document}

   \title{Global properties, fractality, and mass segregation in single, paired, and grouped open clusters }
\titlerunning{Global properties, fractality, and mass segregation in single, paired, and grouped OCs }


   \author{Coenda, V. \inst{1,2}\orcidlink{0000-0001-5262-3822} 
          \and
           Baume, G.\inst{3,4}\orcidlink{0000-0002-0114-0502}
          \and
          Palma, T.\inst{2,5}\orcidlink{0000-0002-0732-2737}
          \and
          Feinstein, C.\inst{3,4}\orcidlink{0000-0003-2341-0494}
          }

     \institute
   {
    Instituto de Astronom\'ia Te\'orica y Experimental (IATE), CONICET, Universidad Nacional de C\'ordoba, Laprida 854, X5000BGR, C\'ordoba, Argentina
    \and    
    Observatorio Astronómico, Universidad Nacional de Córdoba, Laprida 854, X5000BGR, Córdoba, Argentina
    \and
    Facultad de Ciencias Astronómicas y Geofísicas, Universidad Nacional de La Plata, Argentina  
    \and
    Instituto de Astrofísica de La Plata (IALP), CONICET, Universidad Nacional de La Plata, Argentina  
    \and
    Consejo Nacional de Investigaciones Cient\'ificas y T\'ecnicas de la Rep\'ublica Argentina (CONICET)
\\
    \email{vcoenda@unc.edu.ar}
   }

   \date{\today}

 
  \abstract
   {Open clusters (OCs) provide key insights into the formation and dynamical evolution of stellar systems. While many studies have focused on individual clusters, the influence of environmental factors on their structural properties remains an open question.}
   {We investigate the structural and dynamical properties of OCs classified as single, in pairs, or in groups. By analysing their mass, size, age, fractality, and mass segregation, we aim to identify systematic differences among these categories and assess the role of the Galactic environment in their evolution.
   }
   {We analysed a sample of 420 single OCs, 415 in pairs, and 317 in groups, with total masses $2.2\le\log(M/M_{\odot})\le 3.7$. To characterise their structure, we applied the $Q$ parameter, which differentiates fractal from radial distributions. Additionally, we calculated the local density ratio to quantify the mass segregation and explore its dependence on cluster environment.}
   {OCs in groups tend to be the youngest, followed by those in pairs and then single OCs. Although sizes are similar, OCs in pairs and groups tend to be less concentrated. Structurally, grouped OCs exhibit the highest fractality, which decreases with age as clusters evolve towards more radial configurations. Mass segregation is observed in 80\% of OCs, with a marginally higher incidence in single clusters. Some older single OCs show inverse segregation, with massive stars at larger radii. Spatially, single OCs are more dispersed, whereas paired and grouped OCs are concentrated in spiral arms and active star-forming regions.}
   {OC evolution appears to be influenced by a combination of internal dynamics and environmental factors. Single OCs tend to exhibit characteristics consistent with more advanced dynamical evolution, whereas those in pairs and groups may retain structural features that reflect their formation environment.
   The presence of substructures and fractality in younger clusters suggests that interactions within their birth environment play a crucial role in shaping their long-term development. More massive OCs evolve towards radial configurations, while less massive ones may retain fractal characteristics for longer periods. These findings support the idea that both intrinsic properties and external environmental factors play a role in shaping the evolution of OCs.}

   \maketitle
  
%

\section{Introduction}\label{sec:intro}

Star clusters serve as essential astrophysical laboratories, providing insights into star formation, stellar evolution, and galactic dynamics. Dense stellar environments, where stars are typically born, dissolve within a few million years, making star clusters the building blocks of galaxies (\citealt{Lada:2003, deGtijs:2007, Parmentier:2008, Martinez_Barbosa:2016}). Formed within giant molecular clouds, open clusters (OCs) can evolve into various structures, including binary and higher multiplicity systems, offering a unique opportunity to study these evolutionary processes \citep{Carmargo:2016}. The study of OC parameters — such as distance, reddening, age, and metallicity — enables detailed investigations that are not possible with field stars \citep{PieckaPaunzen21}. 

Most stars in the Galactic disc are believed to be born in OCs, which eventually dissipate due to relaxation-driven mass loss or tidal perturbations (\citealt{Lada:2003, Krumholz:2019}). The morphology of OCs is significantly influenced by these processes, with initial survival requiring the expulsion of residual gas after star formation \citep{Baumgardt:2007}. Following this gas expulsion, star clusters are not expected to remain in virial equilibrium, leading to a phase of violent relaxation where young stars can become gravitationally unbound and star clusters thus expand (\citealt{Ernst:2015, Dinnbier:2020a, Dinnbier:2020b}). Key factors affecting the degree of expansion and potential disruption of young clusters include star formation efficiency, gas expulsion timescales, and primordial mass segregation (\citealt{Lada:1984, Goodwin:2009,  Brinkmann:2017, Shukirgaliyev:2017, Shukirgaliyev:2018}). 

Binary clusters are particularly valuable as natural laboratories for advancing our understanding of star cluster formation and evolution. As outlined by \citet{delaFuente09}, there are five primary scenarios for the formation of multiple star clusters. The first is simultaneous formation, where OCs in genetic pairs have the same space velocities, ages, and chemical compositions, suggesting a shared origin from the same molecular cloud or the merger of smaller clusters \citep{Priyatikanto16}. The second scenario involves sequential formation, where stellar winds or supernova shocks from one cluster induce the collapse of a nearby cloud, leading to the formation of a companion cluster \citep{Goodwin97}. The third is tidal capture, where star clusters that form independently are later drawn together through gravitational forces \citep{vandenBergh96}. The fourth is resonant trapping, where interactions within the Galaxy's gravitational field lead to the formation of binary clusters \citep{DehnenBinney98}. The final scenario is optical doubles, meaning pairs that appear close in the sky but are not physically associated. 

The dynamical history and future of binary star clusters have been the subject of several numerical investigations. \citet{delaFuente:2010} employed N-body simulations to examine the evolution of binary star clusters, particularly the impact of the Milky Way’s tidal field on these systems. Their research revealed that binary clusters with smaller separations ($\lesssim$ 10 pc) are prone to merging, whereas those with wider separations are more susceptible to disruption by the Galactic tidal forces. \citet{Priyatikanto16}\footnote{The isolated cluster contains 4096 stars whose total mass is 2064 $M_{\odot}$. Apart from this, its initial conditions are identical to those adopted for the binary cluster simulations.} demonstrated that binary clusters, even with identical initial orbital parameters, can have vastly different dynamical outcomes based on their initial orientation and phase within the Galactic orbit. Furthermore, \citet{Darma:2021}, through N-body simulations over a 50 Myr period, discovered that clusters with more pronounced initial substructures are more likely to evolve into binary systems, particularly in the early phases following a period of violent relaxation.

From an observational perspective, various studies have investigated the structural parameters of OCs (e.g. \citealt{Meingast:2019, Hetem19, Jerabkova:2021}). Typically, the morphology of OCs is characterised by a dense central core surrounded by an outer halo, or corona, where the stellar density is considerably lower \citep{Artyukhina:1964}. Research by \citet{Nilakshi:2002}, later corroborated by \citet{Meingast:2021} and \citet{Tarricq:2022}, has shown that these halos are significantly more extended than the cores and are believed to encompass a substantial portion of the cluster's members.  Understanding the spatial distribution and structural parameters of OCs across various ages and environments provides valuable insight into these evolutionary processes, offering new constraints for theoretical models.

Until recently, the Milky Way global survey of star clusters catalogue (MWSC, \citealt{Kharchenko13}) had provided the most homogeneous and largest sample of star clusters, comprising 3006 clusters (including OCs, globular clusters, and compact associations). The latest addition of 139 OCs by \citet{Schmeja14} and 63 clusters by \citet{Scholz15} brings the total sample to 3208 confirmed clusters, which can be used to study mass segregation for a wide range of ages. However, by using the \citet{Hunt2024} catalogue, we benefit from a sample that is more than twice as large, enabling a more statistically significant analysis of cluster properties and offering new insights into the dynamics and evolution of star clusters. Section \ref{sec:sample} describes the OC sample used in this work. In Sect. \ref{sec:analysis} we present the structural and dynamical analysis of the selected objects, and Sect. \ref{sec:conclusions} summarises our conclusions.

\section{The sample of open clusters}\label{sec:sample}

The OC sample analysed in this work was obtained from \citet[hereafter Paper I]{Palma:2025}, which was derived from \citet{Hunt2023} and \citet{Hunt2024}, who conducted a comprehensive search for OCs using the third \textit{Gaia} Data Release \citep{Gaia:dr3}\footnote{\url{https://www.cosmos.esa.int/web/gaia/dr3}}. Their study produced a homogeneous catalogue of 7167 star systems, 79\% of which were classified as OCs, with the remainder identified as moving groups. In \citetalias{Palma:2025}, the star clusters are categorised into groups, pairs, singles, and unclassified systems. The tidal forces that act on each cluster are computed, taking into account only the influence of the nearest neighbour. The tidal force estimation is calculated using the tidal factor as  $d^3/(M_b r_{50})$, where $d$ is the distance to the closest neighbour (in parsecs), $M_b$ is the total mass of the neighbour cluster (in solar masses), and $r_{50}$ is the radius that encompasses 50\% of the cluster members within the tidal radius (in parsecs).  The tidal factor parameter is the reciprocal to the tidal force.

The star cluster classification is based on the following criteria:
\begin{itemize}
\item Groups (G): sets of three or more clusters, each within 50 pc of one another, all exhibiting tidal factor values of less than 200.
\item Pairs (P): two neighbouring clusters less than 50 pc apart, where at least one has a tidal factor value below 200, with no other clusters in proximity.
\item Singles (S): clusters with no neighbours within a 100 pc radius.
\item Unclassified: clusters that do not satisfy the criteria for any of the aforementioned categories.
\end{itemize}

The analysis identified 2052 single star clusters, 1234 star clusters in pairs, and 936 star clusters in groups.
The distribution of member stars is visualised in a multi-dimensional space, integrating spatial coordinates, parallax, proper motions, and the colour-magnitude diagram. Pairs are further classified as genetic binaries (B), if the two clusters form simultaneously; tidal captures (C); and optical binaries (O), where the clusters are not gravitationally bound but appear close in the sky. Optical binaries sharing the same age are classified as Oa. Further details are provided in \citetalias{Palma:2025}.

In this study, we focus exclusively on the sample of OCs within the three possible environments: singles, pairs, and groups. This yields a sample of 1772 singles, 932, and 683 OCs in pairs and groups, respectively. To achieve our objectives, we used the physical properties derived for the complete sample provided by \citet{Hunt2023} and \citet{Hunt2024}, focusing specifically on total mass, size, and age. In summary, the authors determined the total cluster mass $M$ by deriving photometric masses for member stars through PARSEC isochrone fits in the $G$ band \citep{Bressan:2012}, followed by adjustments for selection biases and unresolved binaries. Subsequently, a mass function was fitted to each cluster and integrated to estimate its total mass, employing a \citet{Kroupa:2001} initial mass function and the \texttt{imf} Python package\footnote{\url{https://github.com/keflavich/imf}}. 

\section{Data analysis}\label{sec:analysis}


\begin{figure*}
\centering
\includegraphics[width=\textwidth]{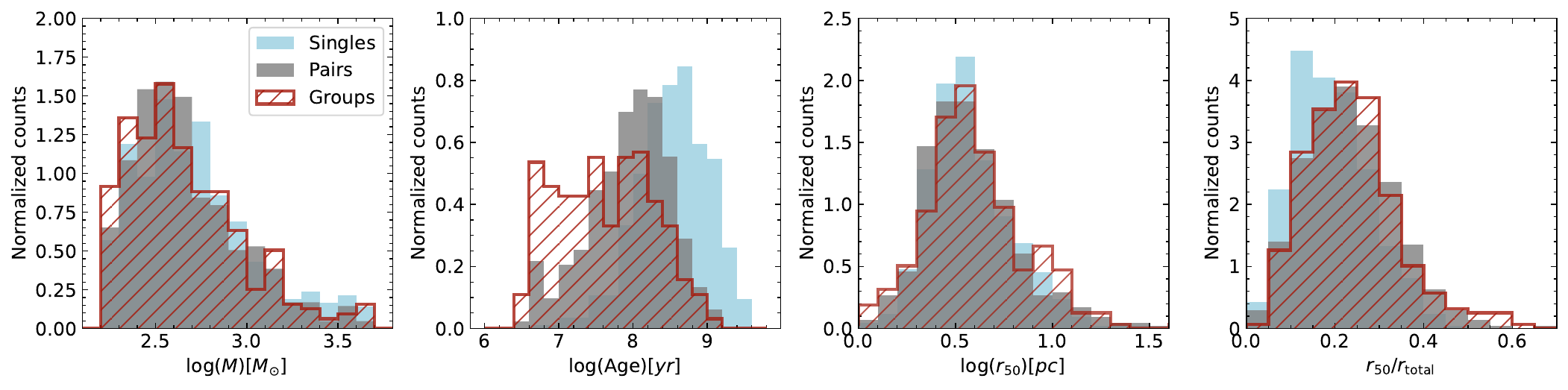}
\caption{Distribution of OCs properties (from left to right): total mass, age, size ($r_{50}$), and concentration ($r_{50}/r_{\rm total}$). Red lines represent OCs in groups, the grey-shaded region corresponds to OCs in pairs, and the light blue region indicates single OCs.}
\label{fig:histo}
\end{figure*}


\begin{figure}
\centering
\includegraphics[width=0.7\columnwidth]{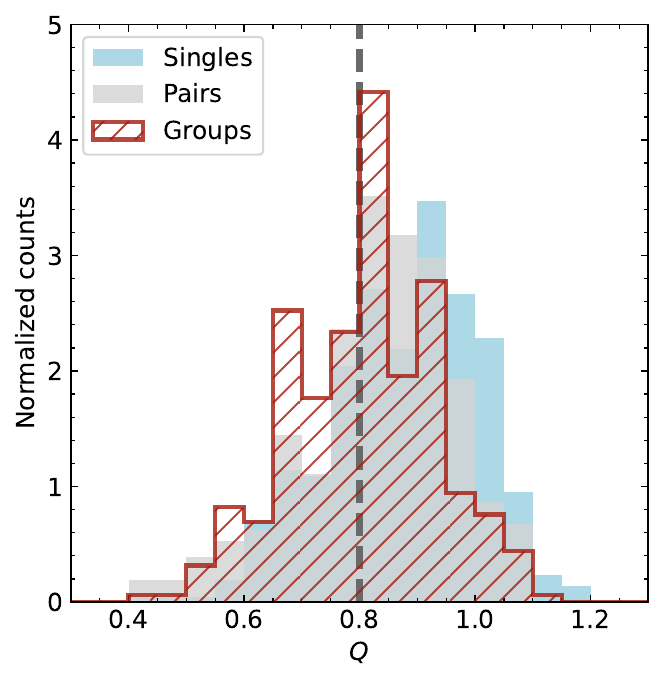}
\caption{Distribution of the OCs' $Q$ parameters. Colours are as in Fig. \ref{fig:histo}. The vertical dashed grey line corresponds to $Q=0.8$, the threshold separating fractal substructures from radial density profiles.}
\label{fig:Q}
\end{figure}


\begin{figure*}
    \centering
    \includegraphics[width=\textwidth]{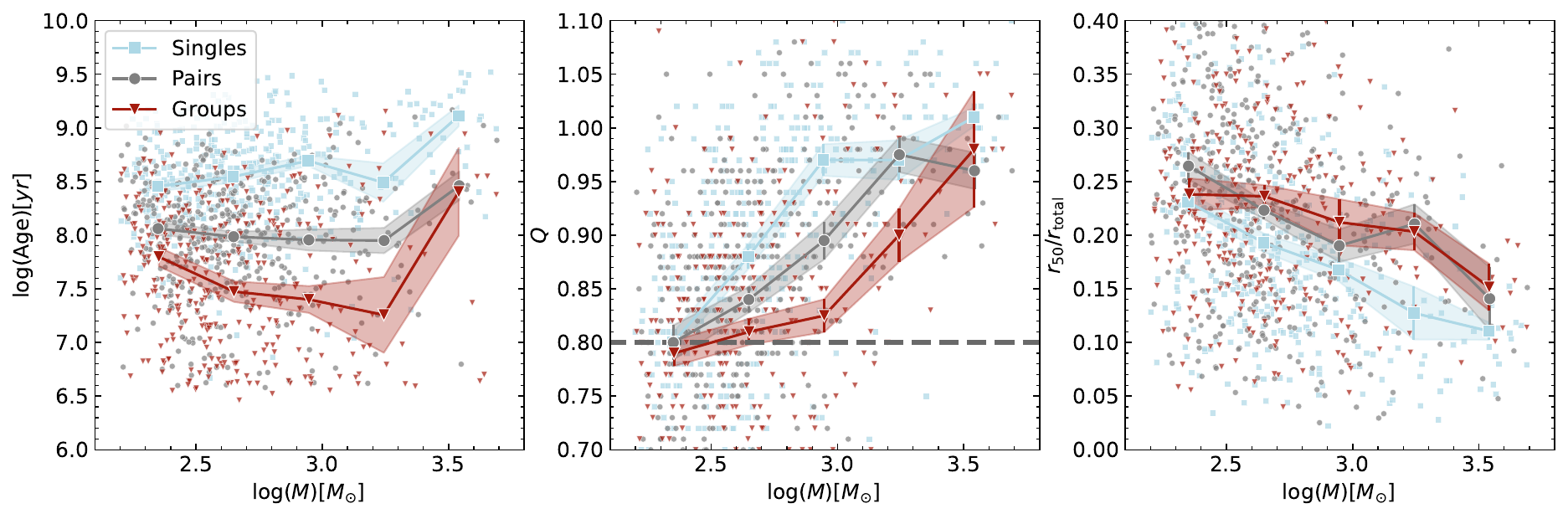}
    \caption{Median properties of OCs as a function of total mass (from left to right): age, $Q$ values, and concentration. Single OCs are represented by light blue squares, pairs by grey circles, and OCs in groups by red triangles. Shaded areas indicate errors obtained via the bootstrap resampling technique. The horizontal dashed grey line in the central panel corresponds to $Q=0.8$}
    \label{fig:props}
\end{figure*}


\begin{figure*}
    \centering
    \includegraphics[width=\textwidth]{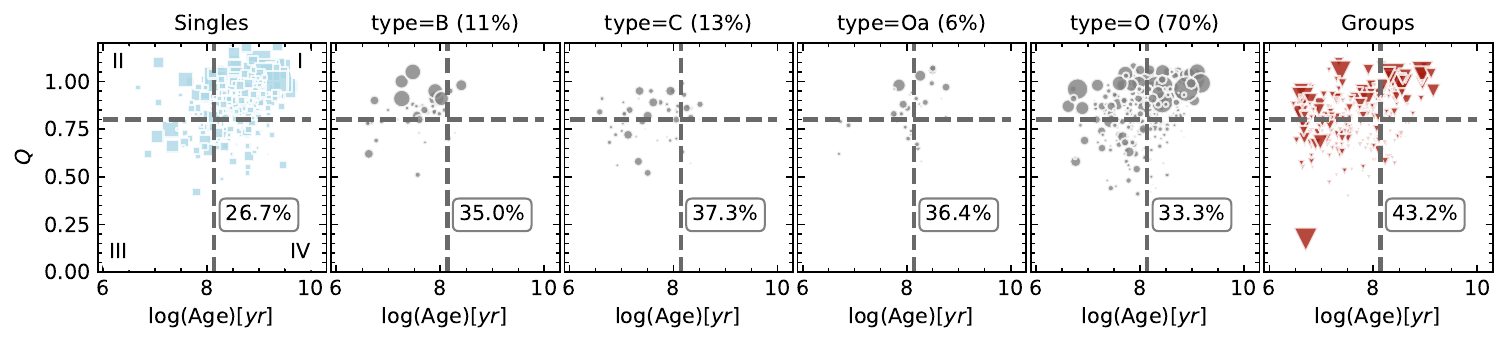}
    \caption{$Q$ parameter as a function of age for (from left to right): singles, pair subtypes B, C, Oa, and O, and groups. The percentages in the panel titles represent the fraction of each pair type relative to the total. The percentages in the boxes indicate the fraction of OCs with $Q<0.8$. Marker sizes correspond to total mass. The horizontal dashed grey line marks $Q=0.8$, while the vertical dashed grey lines indicate the median age of the pairs in the OC system ($\log$(Age)[yr] = $8.1$). In the left panel, we divide the plane into four quadrants, labelled I to IV.}
    \label{fig:pairs}
\end{figure*}

The primary objective of this work is to analyse the scaling relations, internal structure, and mass segregation of OCs, and to compare these properties across diverse environments. For a robust statistical analysis, we restricted the sample to clusters that have more than 30 members, are sufficiently deep ($G \leq 18$) to avoid photometric completeness issues, and with an absolute magnitude in the G band $M_G < +4$. This last limit corresponds approximately to an F8-9V spectral type star. We used clusters distances and reddening values from \citet{Hunt2023} to estimate the absolute magnitudes $M_G$ of each cluster member. 

To minimise mass bias, we included only OCs with total masses in the range $2.2\le\log(M/M_{\odot})\le 3.7$. Within this mass range, our final sample comprises 420 single OCs, 415 pairs, and 317 groups.
Figure \ref{fig:histo} presents the distributions of global parameters (total mass, age, size, and concentration) for our OC sample across different environments (singles, pairs, and groups). In this work, we derived several statistical parameters, reported in Table \ref{tab:stats}, including the median values with their bootstrap uncertainties, the interquartile ranges (25–75), and the 95\% confidence intervals. Furthermore, we computed the 
$p$-values associated with the Kolmogorov–Smirnov (KS), Anderson–Darling (AD), and Cramér–von Mises (CVM) statistical tests, along with the AD test statistic $A^2$; these results are summarised in Table \ref{tab:test}.

Based on the statistical parameters presented in Tables~\ref{tab:stats} and~\ref{tab:test}, we do not find statistically significant differences in the mass distributions among the three groups of OCs (singles, pairs, and groups). The median values and confidence intervals are closely aligned across the samples. Consistently, the statistical tests performed — including the AD, CVM, and KS tests — support the hypothesis that the groups are drawn from similar parent distributions. In particular, all calculated $p$ values are greater than 0.05, and AD test statistics remain below the respective rejection thresholds at the 5\% significance level. We therefore conclude that the mass does not vary significantly between singles, pairs, and groups. 

In contrast to the mass distributions, the age distributions reveal statistically significant differences among the three OC environments. The median values for $\log({\rm Age})\, [{\rm yr}]$ are $8.56\pm0.03$ for single clusters, $8.01\pm0.03$ for pairs, and $7.60\pm0.08$ for groups (see Table~\ref{tab:stats}). These trends are consistent with the confidence intervals 95\%, which do not overlap. Similarly, the interquartile ranges also shift towards younger ages for more dynamically complex environments. The statistical tests summarised in Tables~\ref{tab:stats} and~\ref{tab:test} consistently reject the null hypothesis of a common parent distribution. All $p$ values of the KS, AD, and CVM tests are well below the 0.05 threshold, and the AD statistics exceed the corresponding rejection levels. These results strongly indicate that age distributions differ significantly among the three environments.

On the other hand, the distribution of cluster sizes, measured by $r_{50}$, is statistically indistinguishable between environments. As shown in Tables~\ref{tab:stats} and~\ref{tab:test}, the median sizes and their confidence intervals are remarkably similar for singles, pairs, and groups. Moreover, all tests (KS, AD, and CVM) yield high $p$ values and statistics below their respective rejection thresholds, indicating that the null hypothesis of a common parent distribution cannot be rejected.

We observe that single OCs tend to be less concentrated than those found in pairs or groups. This pattern is apparent in both the median concentration values and the shape of the distributions. Although pairs and groups exhibit very similar concentration profiles, singles are systematically shifted towards lower values.
This behaviour is supported by several statistical tests (AD, CVM, and KS), all of which indicate significant differences between the distributions of singles and the other two categories. In contrast, no significant differences are found between pairs and groups. These results suggest that the presence of neighbouring clusters — either a single companion or several — is associated with higher central concentrations, possibly due to dynamical interactions or common formation environments.

\begin{figure*}
    \centering
    \includegraphics[width=\textwidth]{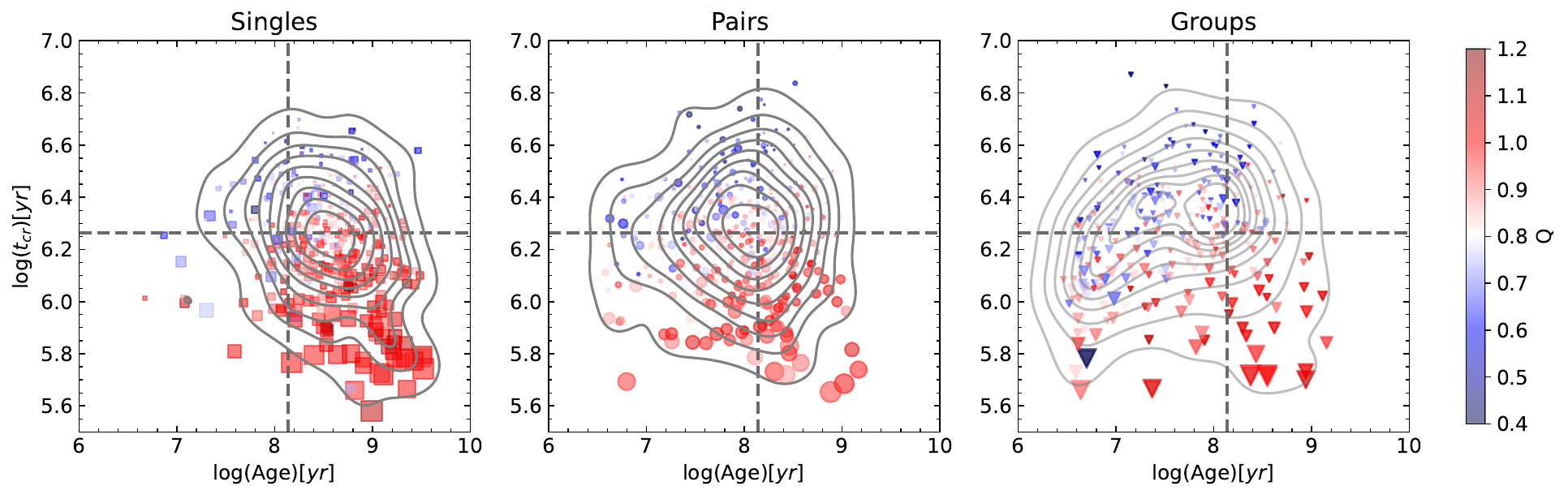}
    \caption{Crossing time ($t_{\rm cr}$) as a function of system age for (from left to right): single OCs, OCs in pairs, and OCs in groups. Symbol sizes corresponds to total mass, and dots are colour-coded by their $Q$ parameter value. Vertical and horizontal dashed grey lines indicate the median values of age and crossing time for the entire sample, respectively. Solid grey lines represent iso-density contours.}
    \label{fig:tiempos}
\end{figure*}

\subsection{Global properties and internal structure of the clusters}

The internal structure of star clusters can be quantified using the $Q$ parameter introduced by \cite{Cartwright04}. This metric is applicable to both observational data and simulations (e.g. \citealt{Delgado:2013, Jaffa:2017, Hetem19, Daffern-Powell:2020}). The $Q$ value helps distinguish between different structural distributions: clusters with $Q < 0.8$ exhibit a fractal structure, while those with $Q > 0.8$ have a radial structure with central concentration. A value of $Q \sim 0.8$ suggests a uniform density profile.

Briefly, the $Q$ parameter is derived from the minimum spanning tree (\citealt{GR1969}) constructed from the projected positions of cluster members. The minimum spanning tree is a unique network that connects all considered stars with straight lines, avoiding loops while minimising the total edge length. Next, the normalised mean separation value of the bound stars ($\bar{m}$), and the normalised mean separation of all stars ($\bar{s}$) are computed. Finally, the $Q$ parameter is given by the ratio $Q = \bar{m}/\bar{s}$. In this work, we computed the $Q$ values considering those stars with a membership probability greater than 50\%. Uncertainties in $\bar{m}$, $\bar{s}$, and $Q$ were estimated using the bootstrap method \citep{efron1992bootstrap}. To prevent computational divergences in the \( Q \) calculations, we relaxed our criterion and included clusters with at least 20 members, following \citep{Hetem19}.

The distributions of the obtained $Q$ values for single, paired, and grouped OCs are shown in Fig.~\ref{fig:Q}. The median value of the $Q$ parameter, shown in in Tables~\ref{tab:stats}, is higher for single clusters and decreases as one moves from pairs to groups. This suggests that, on average, single clusters (which are not in direct interaction) tend to have a higher $Q$ value, possibly indicating greater homogeneity or less influence from interactions. In contrast, pairs and groups exhibit lower $Q$ values, which may reflect a higher degree of substructure or interaction within these systems. The bootstrap error in the median is minimal ($\pm0.01$), indicating that the median estimates are highly precise for each group. Statistical tests, including the AD, CVM, and KS tests, reveal significant differences between the groups (see Table \ref{tab:test}). With these results, the fraction of fractal OCs is 26.7\% for single OCs, 34.2\% for pairs, and 43.2\% for groups.

\begin{figure*}
    \centering
    \includegraphics[width=\textwidth]{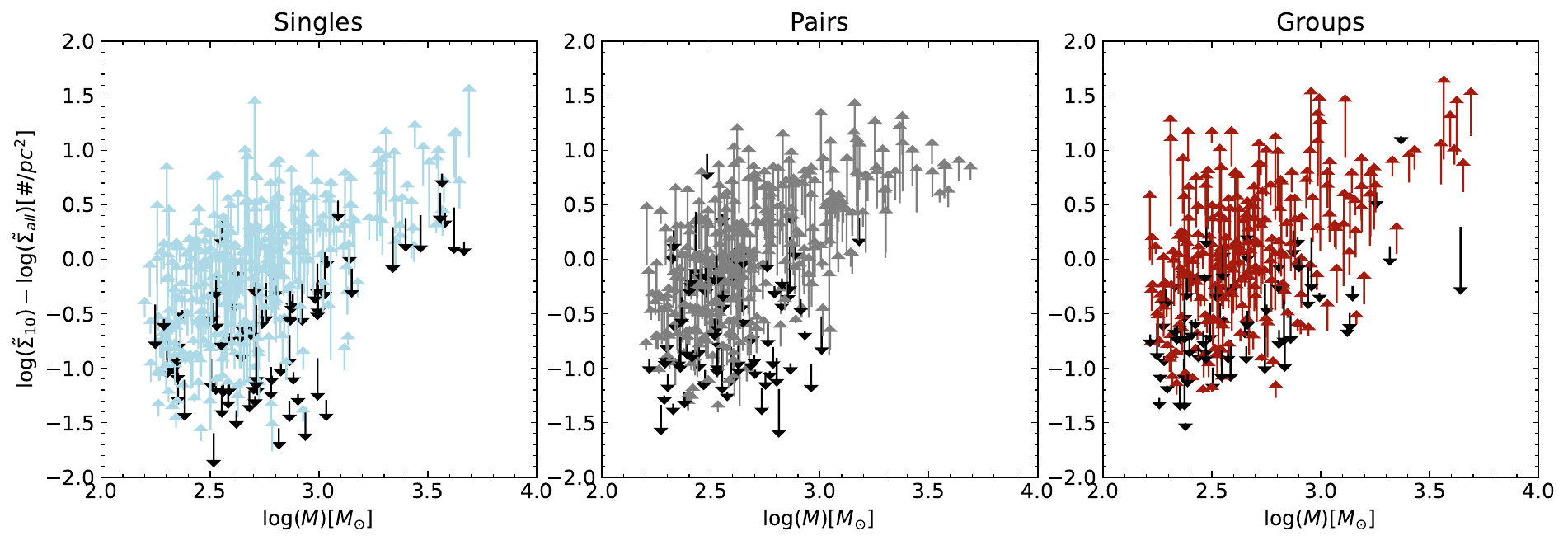}
    \caption{Local density ratio as a function of total mass for (from left to right): single OCs, OCs in pairs, and OCs in groups. Upward arrows indicate OCs with mass segregation, while black downward arrows indicate OCs with inverse mass segregation.}
    \label{fig:LDR}
\end{figure*}

\begin{figure*}
    \centering
    \includegraphics[width=\textwidth]{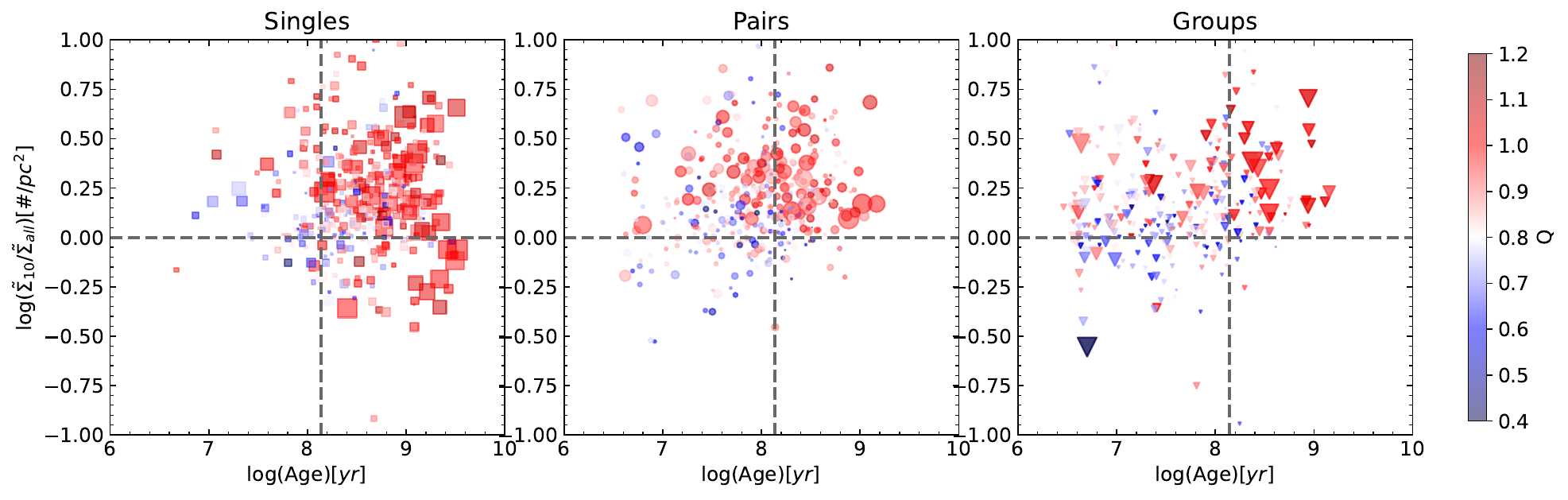}
    \caption{Local density ratio as a function of age. Symbol sizes correspond to total mass, and dots are colour-coded by their $Q$ parameter value. Vertical and horizontal dashed lines indicate median values of age and a zero local density, respectively. }
    \label{fig:LDR_age}
\end{figure*}

The median values of several properties as a function of total system mass are shown in Fig. \ref{fig:props}. The analysis is restricted to bins containing at least ten clusters. In the left panel, the median age of the OCs remains relatively constant with total mass up to $\log(M)[M_{\odot}] \sim 3.2$, beyond which it increases for all samples. Specifically, at a given total mass, single OCs are the oldest, followed by OCs in pairs, and then those in groups. The central panel presents the median $Q$ parameter as a function of total mass, showing a general increase across all samples. For a fixed total mass, single OCs exhibit the highest $Q$ values, followed by OCs in pairs and groups, in that order, until $\log(M)[M_{\odot}] \sim 3.2$.  After that, we do not observe a dependence on the total mass. The right panel displays the median concentration as a function of total mass, indicating that more massive OCs are more concentrated. Single OCs are the most concentrated, while no significant difference is observed between OCs in pairs and groups.

The results presented in Fig.~\ref{fig:props} suggest that single OCs are, on average, older systems, while OCs in pairs — and especially those in groups — are younger for a given total mass. These younger systems exhibit less developed radial density profiles and lower concentrations.
One possibility is that these OCs, located in pairs or groups, formed in highly sub-structured regions, retaining the original geometry of the molecular cloud from which they originated \citep{Elmegreen:2008A, Elmegreen:2018}. This scenario is more likely for younger systems. \citet{Kruijssen:2012} proposed that in dense environments, tidal interactions within the larger birth region of the natal molecular cloud complex could play a dominant role in the cluster dissolution process. As a result, a fractal geometry may arise from tidal interactions between the cluster and the Milky Way (e.g. \citealt{Dalessandro:2015, Yeh:2019, Meingast:2021}). Alternatively, a more fractal geometry could emerge from tidal interactions among the OCs within these paired or grouped systems.

To explore the aforementioned hypothesis, we analysed the relationship between the $Q$ parameter and the age of the OCs. Several studies have investigated the relationship between the ages of different objects and their corresponding $Q$ values, including \cite{SanchezAlfaro10} and \cite{Hetem19} for clusters in the Milky Way, as well as \cite{Gieles2008} and \cite{Bastian2009} for populations in the Magellanic Clouds. Figure \ref{fig:pairs} shows the $Q$ parameter as a function of age and total mass for each pair subtype defined in \citetalias{Palma:2025}, as well as for single OCs and those in groups. 
We considered OCs with $\log({\rm Age)[yr]} > 8.1$ to be old and those with $\log({\rm Age)[yr]}< 8.1$ to be young. 
The adopted limit corresponds to the median age of the sample, irrespective of the environment. 
In general, old massive systems exhibit radial density profiles. 

Among OCs in pairs, approximately $70\%$ of the systems belong to O subtype. The Oa subtype accounts for only $6\%$ of the sample, limiting the analysis due to low statistics. In particular, fractal B and C pairs generally have $\log({\rm Age)[yr]} < 8.1$, a trend that is also observed for OCs in groups. The young age of these types suggests that this morphology reflects the structure of the parent molecular cloud. 

For single OCs, a fractal structure is observed in both young and old systems. Notably, in O-type binary clusters and OCs in groups, older fractal systems tend to be less massive than their younger counterparts, possibly due to stellar mass loss over time, potentially enhanced by tidal interactions. In contrast, single OCs, with intermediate total masses and more isolated environments, may retain both their mass and their initial substructure for longer periods, evolving more slowly towards radial configurations.
The oldest reported clusters with observed spatial structure have ages around $\log({\rm Age)[yr]} \approx 8$ ($100$\,Myr), as reported by \citet{Sanchez:2009}. However, in our sample, we identify fractal objects with ages of $\log({\rm Age)[yr]} = 9.5$, $8.4$, and $8.7$ for single OCs, and those residing in pairs and groups, respectively.

We stress that external tidal interactions do not reduce a cluster’s half-mass radius. In fact, tidal perturbations often stretch a cluster and enhance substructure (leading to a more fractal appearance) without significantly truncating $r_{50}$. Our data show that fractal clusters tend to have larger $r_{50}$ on average, consistent with this idea. The lack of a size difference by environment thus suggests that the dominant tidal influence is the global Galactic field rather than local neighbours. Alternatively, many low-mass clusters may simply retain their primordial fractal structure over a long time (e.g. \citealt{GoodwinWhitworth2004, Kruijssen2012}).

We observe a population of young, massive OCs across all environments, predominantly in pairs and groups, with a $Q$ value greater than 0.8. \citet{Dib:2018}, analysing the OC sample from \citet{Cantat+20}, argue that this idea cannot be sustained. In contrast, \citet{Schmeja:2008} provide evidence in favour of our findings based on numerical smoothed-particle hydrodynamics simulations. Their study suggests that evolution is influenced not only by time but also by the object's mass, which aligns with our results.

\subsection{Dynamical stages}

The dynamics of a stellar system can be characterised by the crossing time ($t_{\rm cr}$), which represents the time required for a star to traverse the system. Following \citet{PZMG10}, we calculated the dynamical timescale ($t_{\rm dyn}$) of a cluster, 
i.e. the timescale over which the system reaches dynamical equilibrium:
\begin{equation}
    t_{\rm dyn}= 2\times10^4\,{\rm yr}\,\Big(\frac{M}{10^6M_{\odot}}\Big)^{-1/2}\,\Big(\frac{r_{vir}}{1 {\rm pc}}\Big)^{3/2}
,\end{equation}
where $M$ is the total mass of the system and $r_{vir}$ is the virial radius. We assumed the crossing time is $t_{\rm cr} = 2.8\,t_{\rm dyn}$ \citep{PZMG10, hetem:2015}. 

Figure \ref{fig:tiempos} shows the crossing time ($t_{\rm cr}$) as a function of system age for single OCs, as well as for OCs in pairs and groups. Each system is represented with a symbol size proportional to its mass and colour-coded according to its $Q$ value, allowing us to examine correlations with mass and structure.  Iso-contour density levels illustrate the distinct distributions of OCs in the parameter space for different environments. Massive single OCs tend to exhibit radial profiles, are older, and have short crossing times, indicating
higher stellar velocities compared to less massive systems. This suggests that they are more dynamically evolved. OCs in pairs and groups, however, occupy a broader region in the $t_{\rm cr}$-age plane. Similar to single OCs, we identify a population of older, non-fractal systems with short crossing times. Additionally, there exists a younger population of OCs in pairs, and particularly in groups, characterised by intermediate to high total masses, radial profiles, and short crossing times. This suggests that despite their young ages, stars in these systems already exhibit high stellar velocities and early dynamical evolution. This population is dynamically linked to the massive systems in the second quadrant of Fig.~\ref{fig:pairs}. The extended distribution of OCs in the $t_{\rm cr}$-age plane reflects the influence of dense and structured environments during formation. This supports the idea that OC evolution depends not only on time but also on their mass and environmental conditions. Less massive systems generally exhibit fractal structures, with longer crossing times and younger ages, indicating that they are the least dynamically evolved. 

Focusing on young OCs in pairs and groups, their low velocity dispersion and young ages suggest that their morphology is unlikely to result from tidal disruption. Instead, it likely reflects the filamentary structure of the parent molecular cloud, which rapidly collapsed to form a coeval population. In contrast, older systems with radial profiles and short crossing times, regardless of their environment, represent dynamically evolved OCs. %

\begin{figure*}
    \centering
    \includegraphics[width=\textwidth]{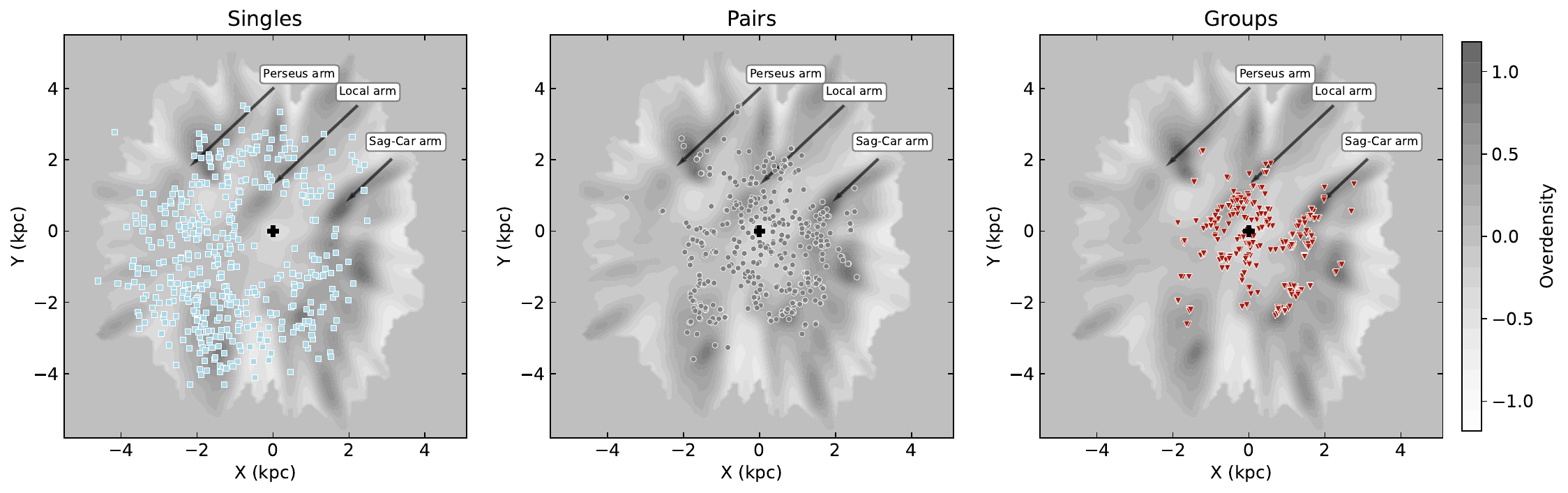}
    \caption{Face-on view of the stellar overdensity map obtained by \citet{Poggio:2021}. The Sun's position is marked with a black cross at $(X, Y) = (0, 0)$. The Galactic centre lies to the right at $(X, Y) = (R_{\odot}=8.2 \,{\rm kpc}, 0)$, with the Galaxy rotating clockwise. The labels indicate the positions of the Perseus arm, the Local (Orion) arm, and the Sagittarius-Carina arm, as defined by \citet{Poggio:2022A}. The panels (from left to right) correspond to single OCs, OCs in pairs, and OCs in groups. Symbols represent the OCs in each environment. }
    \label{fig:overdensity}
\end{figure*}

\subsection{Mass segregation}

Open clusters are typically characterised by mass segregation, a phenomenon in which brighter and more massive stars tend to concentrate towards the cluster's centre, while faint and less massive stars are distributed farther out \citep{Dib:2018, Dib:2019, Hetem19, Maurya:2020, Nony:2021, Hetem:2024}. 
It remains uncertain whether this mass segregation is primarily driven by the cluster's dynamical evolution, the star formation process, or a combination of the two \citep{Dib:2018, Plunkett:2018}.

In the case of dynamical evolution, massive stars transfer their kinetic energy to lower-mass stars during energy equipartition, causing them to accumulate in the cluster’s core, while lower-mass stars move outward at higher velocities \citep{Allison:2009A}. Numerical simulations support this scenario, showing that mass segregation develops over timescales comparable to the cluster's age \citep{Parker:2014}. 

On the other hand, mass segregation can also originate during the star formation process itself, as massive stars preferentially form in the cluster's central regions \citep{Dib:2010}. 
Recent radiation-magneto-hydrodynamic simulations \citep{Guszejnov:2022} suggest that primordial mass segregation occurs during cluster formation. This work reports that small stellar groups in the early stages of formation already exhibit mass segregation, with one or a few massive stars positioned at their centre. As these groups merge, they create mass-segregated substructures within the newly formed cluster. As a result, massive stars are not initially concentrated at the centre of the merged cluster. Over time, dynamical interactions between cluster members lead to a more centralised configuration, resembling that observed in real clusters. This process, known as hierarchical assembly, has been previously discussed in the literature (e.g. \citealt{Bonnel:2003, Vazquez-semadeni2017, Grudic:2018}).

One approach to studying the distribution of massive stars within a star cluster is to calculate their projected local density and compare it to the density of the most massive stars. The projected local density for an individual star $i$ is defined as
\begin{equation}
    \Sigma_i=\frac{N}{\pi r^2_{i,N}}
,\end{equation}
where $r_{i,N}$ represents the distance to the $N^{th}$ nearest neighbour. In this work, we adopted $N=10$. This approach was first applied to galaxy clusters by \citet{Dressler:1980} and later adapted for star clusters by \citet{Maschberger:2011}. \citet{Parker:2014} introduced the local density ratio, defined as 
\begin{equation}
    \Sigma_{LDR}=\frac{\tilde{\Sigma}_{\rm subset}}{\tilde{\Sigma}_{\rm all} }
,\end{equation}
where $\tilde{\Sigma}_{\rm subset}$  is the average local density for a specific subset of stars, and  $\tilde{\Sigma}_{\rm all}$ is the average local density for all stars in the cluster. If the subset contains massive stars located in denser regions, its local surface density will exceed that of the entire sample. In Fig. \ref{fig:LDR} we present $\log(\Sigma_{LDR})=\log(\tilde{\Sigma}_{10})-\log(\tilde{\Sigma}_{\rm all})$ as a function of the total mass for OCs in our sample. To calculate local densities, we included only stars with a membership probability greater than 50\%. We considered a subset of the ten most massive stars. The base of each arrow corresponds to the average local density of the entire sample, while the arrow's tip represents the local density of the ten most massive stars. 
An upward arrow indicates mass segregation (+S), with the most massive stars concentrated towards the cluster centre. In contrast, a downward arrow represents inverse segregation (-S), where these stars are not centrally concentrated.

Regardless of the environment in which the OCs are located, a correlation is evident between the local density ratio and the total mass of the system, with the most massive clusters displaying higher ratios. Furthermore, systems with inverse segregation are observed across the entire mass range. However, the majority of clusters, approximately 80\%, exhibit mass segregation, regardless of their environment. Specifically, inverse segregation is found in  17.9\% of single OCs, 18.3\% of pairs, and 21.4\% of groups. A subtle trend emerges, with the percentage of inverse segregation slightly increasing in systems located in denser environments. 

Figure \ref{fig:LDR_age} shows the local density $\log(\Sigma_{LDR})$ as a  function of the cluster age. As in Fig. \ref{fig:tiempos}, each system is represented with a marker whose size is proportional to its mass and colour-coded according to its $Q$ value.
\citet{Bukowiecki:2012}, based on a sample of 599 OCs from the Two Micron All Sky Survey (2MASS\footnote{https://www.ipac.caltech.edu/2mass/releases/allsky/}), suggested that mass segregation tends to become more pronounced with age. In contrast, \citet{Dib:2018} did not find any noticeable correlation between these parameters. In this regard, we do not observe any clear correlation between local density and age, although the distribution of single OCs in this plane differs from that of OCs in pairs and groups.

As previously mentioned, approximately 80\% of OCs exhibit +S. Among these, 76\% of single OCs, 70\% of paired OCs, and 62\% of group OCs display radial profiles, suggesting that more fractal OCs with +S are found in pairs and groups. Furthermore, 28\% of these objects are young and singles, while 69\% and 80\% are young in pairs and groups, respectively. The high fraction of young and fractal objects — likely unevolved systems still retaining imprints of their parent cloud (III quadrant) — with +S may indicate that a significant fraction of OCs are born segregated, supporting previous findings (e.g. \citealt{Lada:1984, Goodwin:2009, Brinkmann:2017, Shukirgaliyev:2017, Shukirgaliyev:2018}).

We also identify a population of massive OCs with ages $\log({\rm Age}) \geq 8\,  (\sim100$ Myr) that have radial profiles and where there is inverse segregation in single OCs. These objects are located in the first quadrant of Fig.~\ref{fig:pairs}, indicating that they are evolved systems. Examining their stellar distribution, we indeed observe a radial configuration, with many massive stars concentrated in the outer regions. This could be a consequence of an initial spatial substructure, which has been observed in clusters as old as $100$ Myr (e.g. \citealt{Sanchez:2009}). Through dynamical evolution, stars gradually settle into a centralised configuration, with the most massive stars being the last to reach this state or potentially being ejected, as suggested by previous observational studies (e.g. \citealt{Oh:2016, Lennon:2018, Zeidler:2021}).

In pairs and groups, OCs with -S are predominantly young and fractal, comprising 73\% and 88\% of these systems, respectively. These objects retain imprints of their natal cloud and exhibit signs of an initial substructure, which is even more pronounced when they form in groups.


\subsection{Distribution of OCs in overdensity maps} 

Figure \ref{fig:overdensity} displays the overdensity measured by \citet{Poggio:2021}\footnote{\url{https://github.com/epoggio/Spiral_arms_EDR3.git.}} in the Galactic disc's $X - Y$ plane. This map is based on data from upper main sequence stars provided by 2MASS \citep{Skrutskie:2006} and \textit{Gaia} Early Data Release 3 (EDR3) catalogues. The figure highlights the positions of three spiral arm segments near the Sun, which is located at $(X, Y) = (0, 0)\, \Kpc$. We also include the OCs in this map, distinguishing them by their environments.

We observe distinct distributions of OCs depending on whether they reside in groups, pairs, or are singles. Single OCs are typically located at distances $\gtrsim 1$ Kpc from the Sun, with an extended distribution spanning both low- and high-overdensity regions of the Milky Way. OCs in pairs are observed at closer distances and are also found in both density regimes. On the other hand, OCs in groups are more concentrated around the Sun's position and in well-known star-forming regions within the spiral arms. Examples include the Cygnus region and the Orion Nebula in the local arm; the W3/4/5 complex in the Perseus arm; and the W51 giant molecular cloud and the Carina Nebula in the Sagittarius-Carina arm (see \citealt{Reid2019}). Regarding the concentration of pairs and groups of clusters around the Sun, it can be explained by a selection effect due to the uneven behaviour of the errors of the adopted cluster distances. That is, although spiral arms are more likely to exhibit a higher degree of substructure, as evidenced by the presence of OCs in groups and pairs, the growth of uncertainties in their positions with the distance makes it difficult to identify pairs and groups as they are farther from the Sun. This effect is clearly visible in the distributions presented in Fig.~\ref{fig:dist}, since the distributions of OCs identified as pairs or groups drop abruptly when the errors in their distances are $\sim 10-20$ pc, which corresponds to distances $\sim 1.5-2.5$ kpc. 

Figure \ref{fig:mapa2} shows the distribution of OCs in the galactic plane for the four quadrants, as defined in Fig. \ref{fig:pairs}. We divide the sample of OCs in each environment based on their mass segregation and can observe both mass segregation tendencies.  
As discussed in the previous section, we identify a population of massive and old single OCs with inverse segregation in the first quadrant. This population is absent in the other quadrants, where lower-mass objects dominate. In general, non-segregated clusters are located in the densest regions of the overdensity map, associated with the Galactic arms, where substructures are likely to be more prevalent.

\section{Conclusions}\label{sec:conclusions}

In this paper we have compared the properties of OCs classified as group, pair, or single systems based on the sample from \citet{Palma:2025}. Our final dataset comprises 420 single OCs, 415 in pairs, and 317 in groups. To address our objectives, we utilised the physical properties derived for the entire sample by \citet{Hunt2023} and \citet{Hunt2024}, focusing specifically on total mass, size, and age. To analyse the structural properties of these OCs, we employed the $Q$ parameter, which allows us to determine whether their stellar distribution is fractal $(Q < 0.8)$ or radial $(Q > 0.8)$. Additionally, we calculated the local density ratio to investigate mass segregation.
Our key findings are as follows:
\begin{itemize}

\item Age and structure: Grouped OCs tend to be the youngest on average, followed by paired OCs and then single OCs, albeit with overlapping age distributions among the categories. Although no significant differences in size are found, OCs in pairs and groups are less concentrated. The structural differences suggest that single OCs are dynamically more evolved, whereas OCs in pairs and groups retain imprints of their initial formation.

\item Fractality and radial distribution: OCs in groups exhibit the highest degree of fractality, followed by paired OCs, while single OCs are more radially concentrated. The dependence of the $Q$ parameter on total mass is only observed up to  $\log(M)[M_{\odot}]\sim3.2$. Beyond this threshold, no clear correlation with mass is found. The fractal nature of younger OCs suggests that they preserve the structural imprint of their progenitor molecular clouds. In contrast, single OCs can maintain a fractal structure even at older ages, up to $log({\rm Age}) = 9.5$ yr. More massive OCs tend to develop radial structures over time, indicating that dynamical evolution is influenced not only by age but also by mass.

\item Mass segregation: Regardless of the environment, approximately 80\% of OCs exhibit mass segregation. Among these, most have radial profiles: 76\% of single OCs, 70\% in pairs, and 62\% in groups. 
These findings are consistent with the possibility that many OCs exhibit primordial mass segregation, though dynamical evolution may also contribute to the observed distributions.
Some older single OCs exhibit inverse mass segregation, with massive stars located at larger radii, possibly due to dynamical evolution and initial spatial substructures. In pairs and groups, OCs with inverse segregation are predominantly young, further suggesting that they reflect the structure of their natal clouds.
\end{itemize} 

The distribution of OCs in the Galactic plane and its relationship with stellar overdensities and spiral arms was analysed:  
\begin{itemize}
    \item Single OCs tend to be more dispersed and located farther from the Sun, often in lower stellar density regions. This may be due to uncertainties in the distance determination.  
    \item OCs in pairs and groups are more concentrated in higher-density regions and spiral arms, particularly in active star-forming regions such as Orion, Carina, and Cygnus. 
    \item The lack of mass segregation observed in some OCs is more common in denser regions, indicating a connection between the Galactic environment and cluster structure.
\end{itemize}

These results are consistent with the notion that both internal cluster characteristics and external environmental conditions contribute to the evolutionary processes of OCs.
In single OCs, the absence of mass segregation in some cases could result from dynamical evolution, leading to the redistribution of stars over time. In contrast, the fractality observed in OCs within pairs and groups suggests that their evolution is strongly linked to the initial substructure and dynamics of their star-forming region. We find that more massive OCs tend to evolve towards more concentrated and radial configurations, whereas less massive clusters may retain their fractality for a longer period.

In summary, our results highlight that OCs do not evolve in isolation. While single OCs exhibit signs of advanced dynamical evolution, those in pairs and groups retain structural imprints of their formation, suggesting that interactions with their environment play a crucial role in shaping their long-term development.

\begin{acknowledgements}
We gratefully acknowledge financial support from the Argentinian institutions. Consejo Nacional de Investigaciones Cient\'ificas y T\'ecnicas (CONICET; PIP-2022-11220210100064CO and PIP-2022-11220210100714CO), Agencia Nacional de Promoción de la Investigación, el Desarrollo Tecnológico y la Innovación (PICT-2020-3690), Secretaría de Ciencia y Tecnología de la Universidad Nacional de Córdoba (SECYT-UNC, Res. 258/53) and from the Universidad Nacional de La Plata (project: 11/G182).
\end{acknowledgements}

\bibliographystyle{aa} 

\begin{thebibliography}{79}
\expandafter\ifx\csname natexlab\endcsname\relax\def\natexlab#1{#1}\fi

\bibitem[{{Allison} {et~al.}(2009){Allison}, {Goodwin}, {Parker}, {Portegies
  Zwart}, {de Grijs}, \& {Kouwenhoven}}]{Allison:2009A}
{Allison}, R.~J., {Goodwin}, S.~P., {Parker}, R.~J., {et~al.} 2009, \mnras,
  395, 1449

\bibitem[{{Artyukhina} \& {Kholopov}(1964)}]{Artyukhina:1964}
{Artyukhina}, N.~M. \& {Kholopov}, P.~N. 1964, \sovast, 7, 840

\bibitem[{{Bastian} {et~al.}(2009){Bastian}, {Gieles}, {Ercolano}, \&
  {Gutermuth}}]{Bastian2009}
{Bastian}, N., {Gieles}, M., {Ercolano}, B., \& {Gutermuth}, R. 2009, \mnras,
  392, 868

\bibitem[{{Baumgardt} \& {Kroupa}(2007)}]{Baumgardt:2007}
{Baumgardt}, H. \& {Kroupa}, P. 2007, \mnras, 380, 1589

\bibitem[\protect\citeauthoryear{Bonnell, Bate, \& Vine}{2003}]{Bonnel:2003} Bonnell I.~A., Bate M.~R., Vine S.~G., 2003, MNRAS, 343, 413

\bibitem[{{Bressan} {et~al.}(2012){Bressan}, {Marigo}, {Girardi}, {Salasnich},
  {Dal Cero}, {Rubele}, \& {Nanni}}]{Bressan:2012}
{Bressan}, A., {Marigo}, P., {Girardi}, L., {et~al.} 2012, \mnras, 427, 127

\bibitem[{{Brinkmann} {et~al.}(2017){Brinkmann}, {Banerjee}, {Motwani}, \&
  {Kroupa}}]{Brinkmann:2017}
{Brinkmann}, N., {Banerjee}, S., {Motwani}, B., \& {Kroupa}, P. 2017, \aap,
  600, A49

\bibitem[{{Bukowiecki} {et~al.}(2012){Bukowiecki}, {Maciejewski}, {Konorski},
  \& {Niedzielski}}]{Bukowiecki:2012}
{Bukowiecki}, {\L}., {Maciejewski}, G., {Konorski}, P., \& {Niedzielski}, A.
  2012, \actaa, 62, 281

\bibitem[{{Camargo} {et~al.}(2016){Camargo}, {Bica}, \&
  {Bonatto}}]{Carmargo:2016}
{Camargo}, D., {Bica}, E., \& {Bonatto}, C. 2016, \mnras, 455, 3126

\bibitem[{{Cantat-Gaudin} {et~al.}(2020){Cantat-Gaudin}, {Anders},
  {Castro-Ginard}, {Jordi}, {Romero-G{\'o}mez}, {Soubiran}, {Casamiquela},
  {Tarricq}, {Moitinho}, {Vallenari}, {Bragaglia}, {Krone-Martins}, \&
  {Kounkel}}]{Cantat+20}
{Cantat-Gaudin}, T., {Anders}, F., {Castro-Ginard}, A., {et~al.} 2020, \aap,
  640, A1

\bibitem[{{Cartwright} \& {Whitworth}(2004)}]{Cartwright04}
{Cartwright}, A. \& {Whitworth}, A.~P. 2004, \mnras, 348, 589

\bibitem[{{Daffern-Powell} \& {Parker}(2020)}]{Daffern-Powell:2020}
{Daffern-Powell}, E.~C. \& {Parker}, R.~J. 2020, \mnras, 493, 4925

\bibitem[{{Dalessandro} {et~al.}(2015){Dalessandro}, {Miocchi}, {Carraro},
  {J{\'\i}lkov{\'a}}, \& {Moitinho}}]{Dalessandro:2015}
{Dalessandro}, E., {Miocchi}, P., {Carraro}, G., {J{\'\i}lkov{\'a}}, L., \&
  {Moitinho}, A. 2015, \mnras, 449, 1811

\bibitem[{{Darma} {et~al.}(2021){Darma}, {Arifyanto}, \&
  {Kouwenhoven}}]{Darma:2021}
{Darma}, R., {Arifyanto}, M.~I., \& {Kouwenhoven}, M.~B.~N. 2021, \mnras, 506,
  4603

\bibitem[{{de Grijs} \& {Parmentier}(2007)}]{deGtijs:2007}
{de Grijs}, R. \& {Parmentier}, G. 2007, \cjaa, 7, 155

\bibitem[{{de La Fuente Marcos} \& {de La Fuente Marcos}(2009)}]{delaFuente09}
{de La Fuente Marcos}, R. \& {de La Fuente Marcos}, C. 2009, \aap, 500, L13

\bibitem[{{de la Fuente Marcos} \& {de la Fuente
  Marcos}(2010)}]{delaFuente:2010}
{de la Fuente Marcos}, R. \& {de la Fuente Marcos}, C. 2010, \apj, 719, 104

\bibitem[{{Dehnen} \& {Binney}(1998)}]{DehnenBinney98}
{Dehnen}, W. \& {Binney}, J.~J. 1998, \mnras, 298, 387

\bibitem[{{Delgado} {et~al.}(2013){Delgado}, {Djupvik}, {Costado}, \&
  {Alfaro}}]{Delgado:2013}
{Delgado}, A.~J., {Djupvik}, A.~A., {Costado}, M.~T., \& {Alfaro}, E.~J. 2013,
  \mnras, 435, 429

\bibitem[{{Dib} \& {Henning}(2019)}]{Dib:2019}
{Dib}, S. \& {Henning}, T. 2019, \aap, 629, A135

\bibitem[{{Dib} {et~al.}(2018){Dib}, {Schmeja}, \& {Parker}}]{Dib:2018}
{Dib}, S., {Schmeja}, S., \& {Parker}, R.~J. 2018, \mnras, 473, 849

\bibitem[{{Dib} {et~al.}(2010){Dib}, {Shadmehri}, {Padoan}, {Maheswar}, {Ojha},
  \& {Khajenabi}}]{Dib:2010}
{Dib}, S., {Shadmehri}, M., {Padoan}, P., {et~al.} 2010, \mnras, 405, 401

\bibitem[{{Dinnbier} \& {Kroupa}(2020{\natexlab{a}})}]{Dinnbier:2020a}
{Dinnbier}, F. \& {Kroupa}, P. 2020{\natexlab{a}}, \aap, 640, A84

\bibitem[{{Dinnbier} \& {Kroupa}(2020{\natexlab{b}})}]{Dinnbier:2020b}
{Dinnbier}, F. \& {Kroupa}, P. 2020{\natexlab{b}}, \aap, 640, A85

\bibitem[{{Dowd}(2020)}]{twosamples}
{Dowd}, C. 2020, arXiv e-prints, arXiv:2007.01360

\bibitem[{{Dressler}(1980)}]{Dressler:1980}
{Dressler}, A. 1980, \apjs, 42, 565

\bibitem[{Efron(1992)}]{efron1992bootstrap}
Efron, B. 1992, in Breakthroughs in statistics (Springer), 569--593

\bibitem[{{Elmegreen}(2008)}]{Elmegreen:2008A}
{Elmegreen}, B.~G. 2008, \apj, 672, 1006

\bibitem[{{Elmegreen}(2018)}]{Elmegreen:2018}
{Elmegreen}, B.~G. 2018, \apj, 853, 88

\bibitem[{{Ernst} {et~al.}(2015){Ernst}, {Berczik}, {Just}, \&
  {Noel}}]{Ernst:2015}
{Ernst}, A., {Berczik}, P., {Just}, A., \& {Noel}, T. 2015, Astronomische
  Nachrichten, 336, 577

\bibitem[{{Gaia Collaboration} {et~al.}(2023){Gaia Collaboration}, {Vallenari},
  {Brown}, {Prusti}, {de Bruijne}, {Arenou}, {Babusiaux}, {Biermann},
  {Creevey}, {Ducourant}, {Evans}, {Eyer}, {Guerra}, {Hutton}, {Jordi},
  {Klioner}, {Lammers}, {Lindegren}, {Luri}, {Mignard}, {Panem}, {Pourbaix},
  {Randich}, {Sartoretti}, {Soubiran}, {Tanga}, {Walton}, {Bailer-Jones},
  {Bastian}, {Drimmel}, {Jansen}, {Katz}, {Lattanzi}, {van Leeuwen}, {Bakker},
  {Cacciari}, {Casta{\~n}eda}, {De Angeli}, {Fabricius}, {Fouesneau},
  {Fr{\'e}mat}, {Galluccio}, {Guerrier}, {Heiter}, {Masana}, {Messineo},
  {Mowlavi}, {Nicolas}, {Nienartowicz}, {Pailler}, {Panuzzo}, {Riclet}, {Roux},
  {Seabroke}, {Sordo}, {Th{\'e}venin}, {Gracia-Abril}, {Portell}, {Teyssier},
  {Altmann}, {Andrae}, {Audard}, {Bellas-Velidis}, {Benson}, {Berthier},
  {Blomme}, {Burgess}, {Busonero}, {Busso}, {C{\'a}novas}, {Carry}, {Cellino},
  {Cheek}, {Clementini}, {Damerdji}, {Davidson}, {de Teodoro}, {Nu{\~n}ez
  Campos}, {Delchambre}, {Dell'Oro}, {Esquej}, {Fern{\'a}ndez-Hern{\'a}ndez},
  {Fraile}, {Garabato}, {Garc{\'\i}a-Lario}, {Gosset}, {Haigron}, {Halbwachs},
  {Hambly}, {Harrison}, {Hern{\'a}ndez}, {Hestroffer}, {Hodgkin}, {Holl},
  {Jan{\ss}en}, {Jevardat de Fombelle}, {Jordan}, {Krone-Martins}, {Lanzafame},
  {L{\"o}ffler}, {Marchal}, {Marrese}, {Moitinho}, {Muinonen}, {Osborne},
  {Pancino}, {Pauwels}, {Recio-Blanco}, {Reyl{\'e}}, {Riello}, {Rimoldini},
  {Roegiers}, {Rybizki}, {Sarro}, {Siopis}, {Smith}, {Sozzetti}, {Utrilla},
  {van Leeuwen}, {Abbas}, {{\'A}brah{\'a}m}, {Abreu Aramburu}, {Aerts},
  {Aguado}, {Ajaj}, {Aldea-Montero}, {Altavilla}, {{\'A}lvarez}, {Alves},
  {Anders}, {Anderson}, {Anglada Varela}, {Antoja}, {Baines}, {Baker},
  {Balaguer-N{\'u}{\~n}ez}, {Balbinot}, {Balog}, {Barache}, {Barbato},
  {Barros}, {Barstow}, {Bartolom{\'e}}, {Bassilana}, {Bauchet}, {Becciani},
  {Bellazzini}, {Berihuete}, {Bernet}, {Bertone}, {Bianchi}, {Binnenfeld},
  {Blanco-Cuaresma}, {Blazere}, {Boch}, {Bombrun}, {Bossini}, {Bouquillon},
  {Bragaglia}, {Bramante}, {Breedt}, {Bressan}, {Brouillet}, {Brugaletta},
  {Bucciarelli}, {Burlacu}, {Butkevich}, {Buzzi}, {Caffau}, {Cancelliere},
  {Cantat-Gaudin}, {Carballo}, {Carlucci}, {Carnerero}, {Carrasco},
  {Casamiquela}, {Castellani}, {Castro-Ginard}, {Chaoul}, {Charlot}, {Chemin},
  {Chiaramida}, {Chiavassa}, {Chornay}, {Comoretto}, {Contursi}, {Cooper},
  {Cornez}, {Cowell}, {Crifo}, {Cropper}, {Crosta}, {Crowley}, {Dafonte},
  {Dapergolas}, {David}, {David}, {de Laverny}, {De Luise}, {De March}, {De
  Ridder}, {de Souza}, {de Torres}, {del Peloso}, {del Pozo}, {Delbo},
  {Delgado}, {Delisle}, {Demouchy}, {Dharmawardena}, {Di Matteo}, {Diakite},
  {Diener}, {Distefano}, {Dolding}, {Edvardsson}, {Enke}, {Fabre}, {Fabrizio},
  {Faigler}, {Fedorets}, {Fernique}, {Fienga}, {Figueras}, {Fournier},
  {Fouron}, {Fragkoudi}, {Gai}, {Garcia-Gutierrez}, {Garcia-Reinaldos},
  {Garc{\'\i}a-Torres}, {Garofalo}, {Gavel}, {Gavras}, {Gerlach}, {Geyer},
  {Giacobbe}, {Gilmore}, {Girona}, {Giuffrida}, {Gomel}, {Gomez},
  {Gonz{\'a}lez-N{\'u}{\~n}ez}, {Gonz{\'a}lez-Santamar{\'\i}a},
  {Gonz{\'a}lez-Vidal}, {Granvik}, {Guillout}, {Guiraud},
  {Guti{\'e}rrez-S{\'a}nchez}, {Guy}, {Hatzidimitriou}, {Hauser}, {Haywood},
  {Helmer}, {Helmi}, {Sarmiento}, {Hidalgo}, {Hilger}, {H{\l}adczuk}, {Hobbs},
  {Holland}, {Huckle}, {Jardine}, {Jasniewicz}, {Jean-Antoine Piccolo},
  {Jim{\'e}nez-Arranz}, {Jorissen}, {Juaristi Campillo}, {Julbe}, {Karbevska},
  {Kervella}, {Khanna}, {Kontizas}, {Kordopatis}, {Korn}, {K{\'o}sp{\'a}l},
  {Kostrzewa-Rutkowska}, {Kruszy{\'n}ska}, {Kun}, {Laizeau}, {Lambert},
  {Lanza}, {Lasne}, {Le Campion}, {Lebreton}, {Lebzelter}, {Leccia}, {Leclerc},
  {Lecoeur-Taibi}, {Liao}, {Licata}, {Lindstr{\o}m}, {Lister}, {Livanou},
  {Lobel}, {Lorca}, {Loup}, {Madrero Pardo}, {Magdaleno Romeo}, {Managau},
  {Mann}, {Manteiga}, {Marchant}, {Marconi}, {Marcos}, {Marcos Santos},
  {Mar{\'\i}n Pina}, {Marinoni}, {Marocco}, {Marshall}, {Martin Polo},
  {Mart{\'\i}n-Fleitas}, {Marton}, {Mary}, {Masip}, {Massari},
  {Mastrobuono-Battisti}, {Mazeh}, {McMillan}, {Messina}, {Michalik}, {Millar},
  {Mints}, {Molina}, {Molinaro}, {Moln{\'a}r}, {Monari}, {Mongui{\'o}},
  {Montegriffo}, {Montero}, {Mor}, {Mora}, {Morbidelli}, {Morel}, {Morris},
  {Muraveva}, {Murphy}, {Musella}, {Nagy}, {Noval}, {Oca{\~n}a}, {Ogden},
  {Ordenovic}, {Osinde}, {Pagani}, {Pagano}, {Palaversa}, {Palicio},
  {Pallas-Quintela}, {Panahi}, {Payne-Wardenaar}, {Pe{\~n}alosa Esteller},
  {Penttil{\"a}}, {Pichon}, {Piersimoni}, {Pineau}, {Plachy}, {Plum}, {Poggio},
  {Pr{\v{s}}a}, {Pulone}, {Racero}, {Ragaini}, {Rainer}, {Raiteri}, {Rambaux},
  {Ramos}, {Ramos-Lerate}, {Re Fiorentin}, {Regibo}, {Richards}, {Rios Diaz},
  {Ripepi}, {Riva}, {Rix}, {Rixon}, {Robichon}, {Robin}, {Robin}, {Roelens},
  {Rogues}, {Rohrbasser}, {Romero-G{\'o}mez}, {Rowell}, {Royer}, {Ruz Mieres},
  {Rybicki}, {Sadowski}, {S{\'a}ez N{\'u}{\~n}ez}, {Sagrist{\`a} Sell{\'e}s},
  {Sahlmann}, {Salguero}, {Samaras}, {Sanchez Gimenez}, {Sanna},
  {Santove{\~n}a}, {Sarasso}, {Schultheis}, {Sciacca}, {Segol}, {Segovia},
  {S{\'e}gransan}, {Semeux}, {Shahaf}, {Siddiqui}, {Siebert}, {Siltala},
  {Silvelo}, {Slezak}, {Slezak}, {Smart}, {Snaith}, {Solano}, {Solitro},
  {Souami}, {Souchay}, {Spagna}, {Spina}, {Spoto}, {Steele},
  {Steidelm{\"u}ller}, {Stephenson}, {S{\"u}veges}, {Surdej}, {Szabados},
  {Szegedi-Elek}, {Taris}, {Taylor}, {Teixeira}, {Tolomei}, {Tonello}, {Torra},
  {Torra}, {Torralba Elipe}, {Trabucchi}, {Tsounis}, {Turon}, {Ulla}, {Unger},
  {Vaillant}, {van Dillen}, {van Reeven}, {Vanel}, {Vecchiato}, {Viala},
  {Vicente}, {Voutsinas}, {Weiler}, {Wevers}, {Wyrzykowski}, {Yoldas}, {Yvard},
  {Zhao}, {Zorec}, {Zucker}, \& {Zwitter}}]{Gaia:dr3}
{Gaia Collaboration}, {Vallenari}, A., {Brown}, A.~G.~A., {et~al.} 2023, \aap,
  674, A1

\bibitem[{{Gieles} {et~al.}(2008){Gieles}, {Bastian}, \&
  {Ercolano}}]{Gieles2008}
{Gieles}, M., {Bastian}, N., \& {Ercolano}, B. 2008, \mnras, 391, L93

\bibitem[{{Goodwin}(1997)}]{Goodwin97}
{Goodwin}, S.~P. 1997, \mnras, 284, 785

\bibitem[Goodwin \& Whitworth(2004)]{GoodwinWhitworth2004}
Goodwin, S.~P., \& Whitworth, A.~P.\ 2004, \aap, 413, 929. doi:10.1051/0004-6361:20031571

\bibitem[{{Goodwin}(2009)}]{Goodwin:2009}
{Goodwin}, S.~P. 2009, \apss, 324, 259

\bibitem[{Gower \& Ross(1969)}]{GR1969}
Gower, J.~C. \& Ross, G. J.~S. 1969, Journal of the Royal Statistical Society.
  Series C (Applied Statistics), 18, 54

\bibitem[{{Gregorio-Hetem} \& {Hetem}(2024)}]{Hetem:2024}
{Gregorio-Hetem}, J. \& {Hetem}, A. 2024, \mnras, 533, 1782

\bibitem[{{Gregorio-Hetem} {et~al.}(2015){Gregorio-Hetem}, {Hetem},
  {Santos-Silva}, \& {Fernandes}}]{hetem:2015}
{Gregorio-Hetem}, J., {Hetem}, A., {Santos-Silva}, T., \& {Fernandes}, B. 2015,
  \mnras, 448, 2504
  
\bibitem[\protect\citeauthoryear{Grudi{\'c} et al.}{2018}]{Grudic:2018} Grudi{\'c} M.~Y., Guszejnov D., Hopkins P.~F., Lamberts A., Boylan-Kolchin M., Murray N., Schmitz D., 2018, MNRAS, 481, 688


\bibitem[{{Guszejnov} {et~al.}(2022){Guszejnov}, {Markey}, {Offner},
  {Grudi{\'c}}, {Faucher-Gigu{\`e}re}, {Rosen}, \& {Hopkins}}]{Guszejnov:2022}
{Guszejnov}, D., {Markey}, C., {Offner}, S. S.~R., {et~al.} 2022, \mnras, 515,
  167

\bibitem[{{Hetem} \& {Gregorio-Hetem}(2019)}]{Hetem19}
{Hetem}, A. \& {Gregorio-Hetem}, J. 2019, \mnras, 490, 2521

\bibitem[{{Hunt} \& {Reffert}(2023)}]{Hunt2023}
{Hunt}, E.~L. \& {Reffert}, S. 2023, \aap, 673, A114


\bibitem[Hunt \& Reffert(2024)]{Hunt2024} Hunt, E.~L. \& Reffert, S.\ 2024, \aap, 686, A42


\bibitem[{{Jaffa} {et~al.}(2017){Jaffa}, {Whitworth}, \& {Lomax}}]{Jaffa:2017}
{Jaffa}, S.~E., {Whitworth}, A.~P., \& {Lomax}, O. 2017, \mnras, 466, 1082

\bibitem[{{Jerabkova} {et~al.}(2021){Jerabkova}, {Boffin}, {Beccari}, {de
  Marchi}, {de Bruijne}, \& {Prusti}}]{Jerabkova:2021}
{Jerabkova}, T., {Boffin}, H. M.~J., {Beccari}, G., {et~al.} 2021, \aap, 647,
  A137

\bibitem[{{Kharchenko} {et~al.}(2013){Kharchenko}, {Piskunov}, {Schilbach},
  {R{\"o}ser}, \& {Scholz}}]{Kharchenko13}
{Kharchenko}, N.~V., {Piskunov}, A.~E., {Schilbach}, E., {R{\"o}ser}, S., \&
  {Scholz}, R.~D. 2013, \aap, 558, A53

\bibitem[{{Kroupa}(2001)}]{Kroupa:2001}
{Kroupa}, P. 2001, \mnras, 322, 231

\bibitem[{{Kruijssen}(2012)}]{Kruijssen:2012}
{Kruijssen}, J.~M.~D. 2012, \mnras, 426, 3008

\bibitem[Kruijssen(2012)]{Kruijssen2012}
Kruijssen, J.~M.~D.\ 2012, \mnras, 426, 3008. doi:10.1111/j.1365-2966.2012.21973.x


\bibitem[{{Krumholz} {et~al.}(2019){Krumholz}, {McKee}, \&
  {Bland-Hawthorn}}]{Krumholz:2019}
{Krumholz}, M.~R., {McKee}, C.~F., \& {Bland-Hawthorn}, J. 2019, \araa, 57, 227


\bibitem[{{Lada} \& {Lada}(2003)}]{Lada:2003}
{Lada}, C.~J. \& {Lada}, E.~A. 2003, \araa, 41, 57

\bibitem[{{Lada} {et~al.}(1984){Lada}, {Margulis}, \& {Dearborn}}]{Lada:1984}
{Lada}, C.~J., {Margulis}, M., \& {Dearborn}, D. 1984, \apj, 285, 141

\bibitem[{{Lennon} {et~al.}(2018){Lennon}, {Evans}, {van der Marel},
  {Anderson}, {Platais}, {Herrero}, {de Mink}, {Sana}, {Sabbi}, {Bedin},
  {Crowther}, {Langer}, {Ramos Lerate}, {del Pino}, {Renzo},
  {Sim{\'o}n-D{\'\i}az}, \& {Schneider}}]{Lennon:2018}
{Lennon}, D.~J., {Evans}, C.~J., {van der Marel}, R.~P., {et~al.} 2018, \aap,
  619, A78

\bibitem[{{Mart{\'\i}nez-Barbosa} {et~al.}(2016){Mart{\'\i}nez-Barbosa},
  {Brown}, {Boekholt}, {Portegies Zwart}, {Antiche}, \&
  {Antoja}}]{Martinez_Barbosa:2016}
{Mart{\'\i}nez-Barbosa}, C.~A., {Brown}, A.~G.~A., {Boekholt}, T., {et~al.}
  2016, \mnras, 457, 1062

\bibitem[{{Maschberger} \& {Clarke}(2011)}]{Maschberger:2011}
{Maschberger}, T. \& {Clarke}, C.~J. 2011, \mnras, 416, 541

\bibitem[{{Maurya} {et~al.}(2020){Maurya}, {Joshi}, \& {Gour}}]{Maurya:2020}
{Maurya}, J., {Joshi}, Y.~C., \& {Gour}, A.~S. 2020, \mnras, 495, 2496

\bibitem[{{Meingast} \& {Alves}(2019)}]{Meingast:2019}
{Meingast}, S. \& {Alves}, J. 2019, \aap, 621, L3

\bibitem[{{Meingast} {et~al.}(2021){Meingast}, {Alves}, \&
  {Rottensteiner}}]{Meingast:2021}
{Meingast}, S., {Alves}, J., \& {Rottensteiner}, A. 2021, \aap, 645, A84

\bibitem[{{Nilakshi} {et~al.}(2002){Nilakshi}, {Sagar}, {Pandey}, \&
  {Mohan}}]{Nilakshi:2002}
{Nilakshi}, {Sagar}, R., {Pandey}, A.~K., \& {Mohan}, V. 2002, \aap, 383, 153

\bibitem[{{Nony} {et~al.}(2021){Nony}, {Robitaille}, {Motte}, {Gonzalez},
  {Joncour}, {Moraux}, {Men'shchikov}, {Didelon}, {Louvet}, {Buckner},
  {Schneider}, {Lumsden}, {Bontemps}, {Pouteau}, {Cunningham}, {Fiorellino},
  {Oudmaijer}, {Andr{\'e}}, \& {Thomasson}}]{Nony:2021}
{Nony}, T., {Robitaille}, J.~F., {Motte}, F., {et~al.} 2021, \aap, 645, A94

\bibitem[{{Oh} \& {Kroupa}(2016)}]{Oh:2016}
{Oh}, S. \& {Kroupa}, P. 2016, \aap, 590, A107

\bibitem[{{Palma} {et~al.}(2025){Palma}, {Coenda}, {Baume}, \&
  {Feinstein}}]{Palma:2025}
{Palma}, T., {Coenda}, V., {Baume}, G., \& {Feinstein}, C. 2025, \aap, 693,
  A218

\bibitem[{{Parker} {et~al.}(2014){Parker}, {Wright}, {Goodwin}, \&
  {Meyer}}]{Parker:2014}
{Parker}, R.~J., {Wright}, N.~J., {Goodwin}, S.~P., \& {Meyer}, M.~R. 2014,
  \mnras, 438, 620

\bibitem[{{Parmentier} \& {de Grijs}(2008)}]{Parmentier:2008}
{Parmentier}, G. \& {de Grijs}, R. 2008, \mnras, 383, 1103

\bibitem[{{Piecka} \& {Paunzen}(2021)}]{PieckaPaunzen21}
{Piecka}, M. \& {Paunzen}, E. 2021, \aap, 649, A54

\bibitem[{{Plunkett} {et~al.}(2018){Plunkett}, {Fern{\'a}ndez-L{\'o}pez},
  {Arce}, {Busquet}, {Mardones}, \& {Dunham}}]{Plunkett:2018}
{Plunkett}, A.~L., {Fern{\'a}ndez-L{\'o}pez}, M., {Arce}, H.~G., {et~al.} 2018,
  \aap, 615, A9

\bibitem[{{Poggio} {et~al.}(2021){Poggio}, {Drimmel}, {Cantat-Gaudin}, {Ramos},
  {Ripepi}, {Zari}, {Andrae}, {Blomme}, {Chemin}, {Clementini}, {Figueras},
  {Fouesneau}, {Fr{\'e}mat}, {Lobel}, {Marshall}, {Muraveva}, \&
  {Romero-G{\'o}mez}}]{Poggio:2021}
{Poggio}, E., {Drimmel}, R., {Cantat-Gaudin}, T., {et~al.} 2021, \aap, 651,
  A104

\bibitem[{{Poggio} {et~al.}(2022){Poggio}, {Recio-Blanco}, {Palicio}, {Re
  Fiorentin}, {de Laverny}, {Drimmel}, {Kordopatis}, {Lattanzi}, {Schultheis},
  {Spagna}, \& {Spitoni}}]{Poggio:2022A}
{Poggio}, E., {Recio-Blanco}, A., {Palicio}, P.~A., {et~al.} 2022, \aap, 666,
  L4

\bibitem[{{Portegies Zwart} {et~al.}(2010){Portegies Zwart}, {McMillan}, \&
  {Gieles}}]{PZMG10}
{Portegies Zwart}, S.~F., {McMillan}, S. L.~W., \& {Gieles}, M. 2010, \araa,
  48, 431

\bibitem[{{Priyatikanto} {et~al.}(2016){Priyatikanto}, {Kouwenhoven},
  {Arifyanto}, {Wulandari}, \& {Siregar}}]{Priyatikanto16}
{Priyatikanto}, R., {Kouwenhoven}, M.~B.~N., {Arifyanto}, M.~I., {Wulandari},
  H.~R.~T., \& {Siregar}, S. 2016, \mnras, 457, 1339

\bibitem[{{Reid} {et~al.}(2019){Reid}, {Menten}, {Brunthaler}, {Zheng}, {Dame},
  {Xu}, {Li}, {Sakai}, {Wu}, {Immer}, {Zhang}, {Sanna}, {Moscadelli}, {Rygl},
  {Bartkiewicz}, {Hu}, {Quiroga-Nu{\~n}ez}, \& {van Langevelde}}]{Reid2019}
{Reid}, M.~J., {Menten}, K.~M., {Brunthaler}, A., {et~al.} 2019, \apj, 885, 131

\bibitem[{{S{\'a}nchez} \& {Alfaro}(2009)}]{Sanchez:2009}
{S{\'a}nchez}, N. \& {Alfaro}, E.~J. 2009, \apj, 696, 2086

\bibitem[{{S{\'a}nchez} \& {Alfaro}(2010)}]{SanchezAlfaro10}
{S{\'a}nchez}, N. \& {Alfaro}, E.~J. 2010, {The fractal spatial dist. of stars
  in open clusters and stellar associations}, Vol.~4, 1--11

\bibitem[{{Schmeja} {et~al.}(2014){Schmeja}, {Kharchenko}, {Piskunov},
  {R{\"o}ser}, {Schilbach}, {Froebrich}, \& {Scholz}}]{Schmeja14}
{Schmeja}, S., {Kharchenko}, N.~V., {Piskunov}, A.~E., {et~al.} 2014, \aap,
  568, A51

\bibitem[{{Schmeja} {et~al.}(2008){Schmeja}, {Kumar}, {Froebrich}, \&
  {Klessen}}]{Schmeja:2008}
{Schmeja}, S., {Kumar}, M.~S.~N., {Froebrich}, D., \& {Klessen}, R.~S. 2008, in
  IAU Symposium, Vol. 246, Dynamical Evolution of Dense Stellar Systems, ed.
  E.~{Vesperini}, M.~{Giersz}, \& A.~{Sills}, 50--54

\bibitem[{{Scholz} {et~al.}(2015){Scholz}, {Kharchenko}, {Piskunov},
  {R{\"o}ser}, \& {Schilbach}}]{Scholz15}
{Scholz}, R.~D., {Kharchenko}, N.~V., {Piskunov}, A.~E., {R{\"o}ser}, S., \&
  {Schilbach}, E. 2015, \aap, 581, A39

\bibitem[{{Shukirgaliyev} {et~al.}(2017){Shukirgaliyev}, {Parmentier},
  {Berczik}, \& {Just}}]{Shukirgaliyev:2017}
{Shukirgaliyev}, B., {Parmentier}, G., {Berczik}, P., \& {Just}, A. 2017, \aap,
  605, A119

\bibitem[{{Shukirgaliyev} {et~al.}(2018){Shukirgaliyev}, {Parmentier}, {Just},
  \& {Berczik}}]{Shukirgaliyev:2018}
{Shukirgaliyev}, B., {Parmentier}, G., {Just}, A., \& {Berczik}, P. 2018, \apj,
  863, 171

\bibitem[{{Skrutskie} {et~al.}(2006){Skrutskie}, {Cutri}, {Stiening},
  {Weinberg}, {Schneider}, {Carpenter}, {Beichman}, {Capps}, {Chester},
  {Elias}, {Huchra}, {Liebert}, {Lonsdale}, {Monet}, {Price}, {Seitzer},
  {Jarrett}, {Kirkpatrick}, {Gizis}, {Howard}, {Evans}, {Fowler}, {Fullmer},
  {Hurt}, {Light}, {Kopan}, {Marsh}, {McCallon}, {Tam}, {Van Dyk}, \&
  {Wheelock}}]{Skrutskie:2006}
{Skrutskie}, M.~F., {Cutri}, R.~M., {Stiening}, R., {et~al.} 2006, \aj, 131,
  1163

\bibitem[{{Tarricq} {et~al.}(2022){Tarricq}, {Soubiran}, {Casamiquela},
  {Castro-Ginard}, {Olivares}, {Miret-Roig}, \& {Galli}}]{Tarricq:2022}
{Tarricq}, Y., {Soubiran}, C., {Casamiquela}, L., {et~al.} 2022, \aap, 659, A59

\bibitem[{{van den Bergh}(1996)}]{vandenBergh96}
{van den Bergh}, S. 1996, \apjl, 471, L31

\bibitem[\protect\citeauthoryear{V{\'a}zquez-Semadeni, Gonz{\'a}lez-Samaniego, \& Col{\'\i}n}{2017}]{Vazquez-semadeni2017} V{\'a}zquez-Semadeni E., Gonz{\'a}lez-Samaniego A., Col{\'\i}n P., 2017, MNRAS, 467, 1313

\bibitem[{{Yeh} {et~al.}(2019){Yeh}, {Carraro}, {Montalto}, \&
  {Seleznev}}]{Yeh:2019}
{Yeh}, F.~C., {Carraro}, G., {Montalto}, M., \& {Seleznev}, A.~F. 2019, \aj,
  157, 115

\bibitem[{{Zeidler} {et~al.}(2021){Zeidler}, {Sabbi}, {Nota}, \&
  {McLeod}}]{Zeidler:2021}
{Zeidler}, P., {Sabbi}, E., {Nota}, A., \& {McLeod}, A.~F. 2021, \aj, 161, 140

\end{thebibliography}

\clearpage
\begin{appendix}
\section{Additional tables}

\begin{table}[h]
\caption{Summary statistics for various parameters of OCs classified as singles, in pairs, and in groups.  }
\label{tab:stats}
    \centering
\begin{tabular}{llccc}
\toprule
    Parameter & Type of OC &      Median &    P25–P75 &      95\% CI    \\
\midrule
$\log(M)\,[M_{\odot}]$      &    Singles & 2.67 ± 0.02 &  2.47–2.88 & 2.63–2.70  \\
                            &      Pairs & 2.61 ± 0.01 &  2.45–2.83 & 2.58–2.63  \\
                            &     Groups & 2.60 ± 0.02 &  2.43–2.83 & 2.55–2.64  \\ \hline
$\log({\rm Age})\, [\rm yr]$&    Singles & 8.56 ± 0.03 &  8.23–8.90 & 8.50–8.89   \\
                            &      Pairs & 8.01 ± 0.03 &  7.61–8.32 & 7.95–8.07  \\
                            &     Groups & 7.60 ± 0.08 &  7.10–8.10 & 7.50–7.77  \\\hline
$\log(r_{50})\, [\rm pc]$   &    Singles & 0.55 ± 0.01 &  0.43–0.70 & 0.52–0.56  \\
                            &      Pairs & 0.55 ± 0.01 &  0.43–0.73 & 0.54–0.60  \\
                            &     Groups & 0.56 ± 0.01 &  0.41–0.72 & 0.55–0.61  \\ \hline
$r_{50}/r_{\rm total}$      &    Singles & 0.188 ± 0.006 &  0.128–0.250 & 0.175–0.199  \\
                            &      Pairs & 0.228 ± 0.004 &  0.161–0.298 & 0.216–0.236  \\
                            &     Groups & 0.227 ± 0.006 &  0.159–0.288 & 0.215–0.242 \\\hline
$Q$ parameter               &    Singles & 0.89 ± 0.01 &  0.79–0.98 & 0.87–0.91 \\
                            &      Pairs & 0.85 ± 0.01 &  0.76–0.93 & 0.83–0.86 \\
                            &     Groups & 0.81 ± 0.01 &  0.72–0.89 & 0.80–0.83 \\
\bottomrule
\end{tabular}
\tablefoot{For each parameter and cluster type, we report the median value, the interquartile range (P25–P75), and the 95\% confidence interval (CI).}
\end{table}

\begin{table}[h]
\caption{Results of the statistical comparisons between different types of OC groupings for the parameter $\log(M)\,[M_{\odot}]$. }
\label{tab:test}
    \centering
\begin{tabular}{llcc}
\toprule
Parameter & Comparison & $p$-values (KS–AD–CVM) & A$^2$ (AD test) \\
\midrule
$\log(M)\,[M_{\odot}]$      & Singles-Pairs   & 0.056-0.061-0.051  &   2.39  \\
                            & Singles-Groups  & 0.056-0.072-0.065  &   2.27  \\
                            & Pairs-Groups    & 0.694-0.760-0.872  &   0.500  \\ \hline
$\log({\rm Age})\, [\rm yr]$& Singles-Pairs   & $p\ll0.05$         &   114.85 \\         
                            & Singles-Groups  & $p\ll0.05$         &   149.50 \\
                            & Pairs-Groups    & $p\ll0.05$         &   27.41\\ \hline
$\log(r_{50})\, [\rm pc]$   & Singles-Pairs   & 0.679-0.474-0.542   &  0.80978 \\
                            & Singles-Groups  & 0.546-0.239-0.208   &  1.2863 \\
                            & Pairs-Groups    & 0.760-0.496-0.436   &  0.78799  \\ \hline
$r_{50}/r_{\rm total}$      & Singles-Pairs   & $p\ll0.05$          &  14.32 \\                
                            & Singles-Groups  & $p\ll0.05$          &  13.77  \\
                            & Pairs-Groups    & 0.963-0.848-0.881   &  0.41 \\ \hline
$Q$ parameter               & Singles-Pairs   & $p\ll0.05$          &  12.25 \\
                            & Singles-Groups  & $p\ll0.05$          &  28.56 \\
                            & Pairs-Groups    & $p\ll0.05$          & 6.926   \\
\bottomrule
\end{tabular}
\tablefoot{For each comparison (Singles–Pairs, Singles–Groups, Pairs–Groups), we report the p-values obtained from the Kolmogorov–Smirnov (KS), Anderson–Darling (AD), and Cramér–von Mises (CVM) tests, listed in that order. We used the \texttt{twosamples} package \citep{twosamples} and \texttt{Cramer} package, available in the R statistical environment. The value of the Anderson–Darling statistic ($A^2$) is also provided, computed using the \texttt{Ksamples} package available also in R environment.}
\end{table}

\clearpage

\section{Additional figure}
\begin{figure}[h]
\centering
\includegraphics[width=0.9\textwidth]{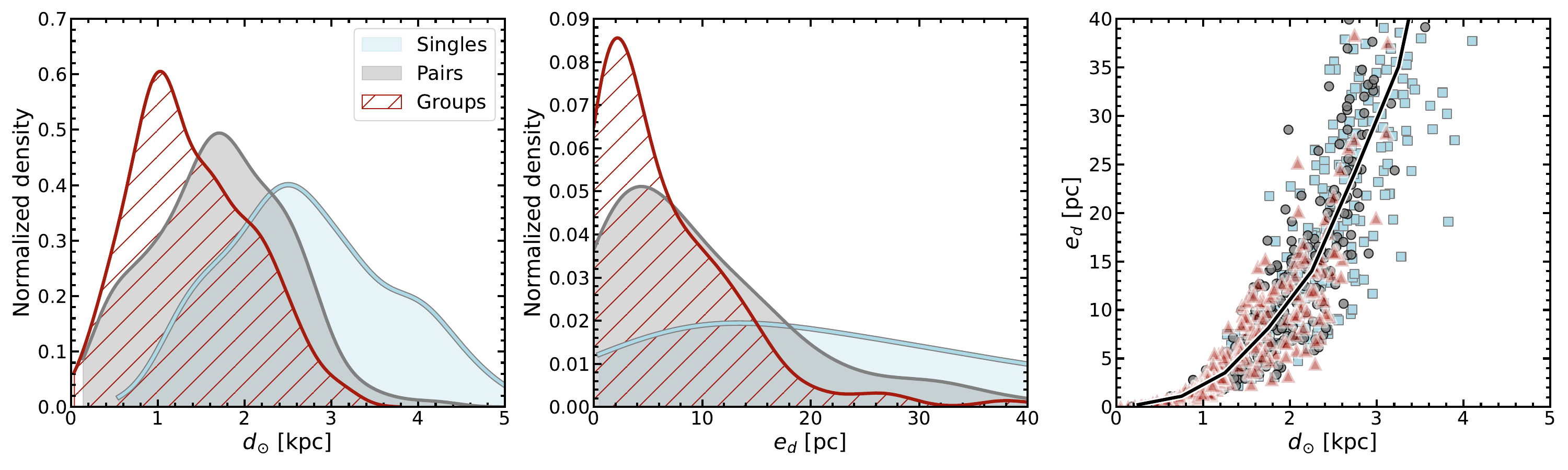}
\caption{Left and central panels: Distributions of OCs' distances to the Sun ($d_{\odot}$) and their errors ($e_d$). Right panel: Relationship between distance and the corresponding error. The black curve indicates the trend of mean values. Colours in all panels follow those in Fig.~\ref{fig:histo}. We estimated $e_d$ using the 16\% and 84\% quantiles of distances from the \citet{Hunt2024} database.}
\label{fig:dist}
\end{figure}

\clearpage
\begin{figure}[h]
\centering
\includegraphics[width=0.9\textwidth]{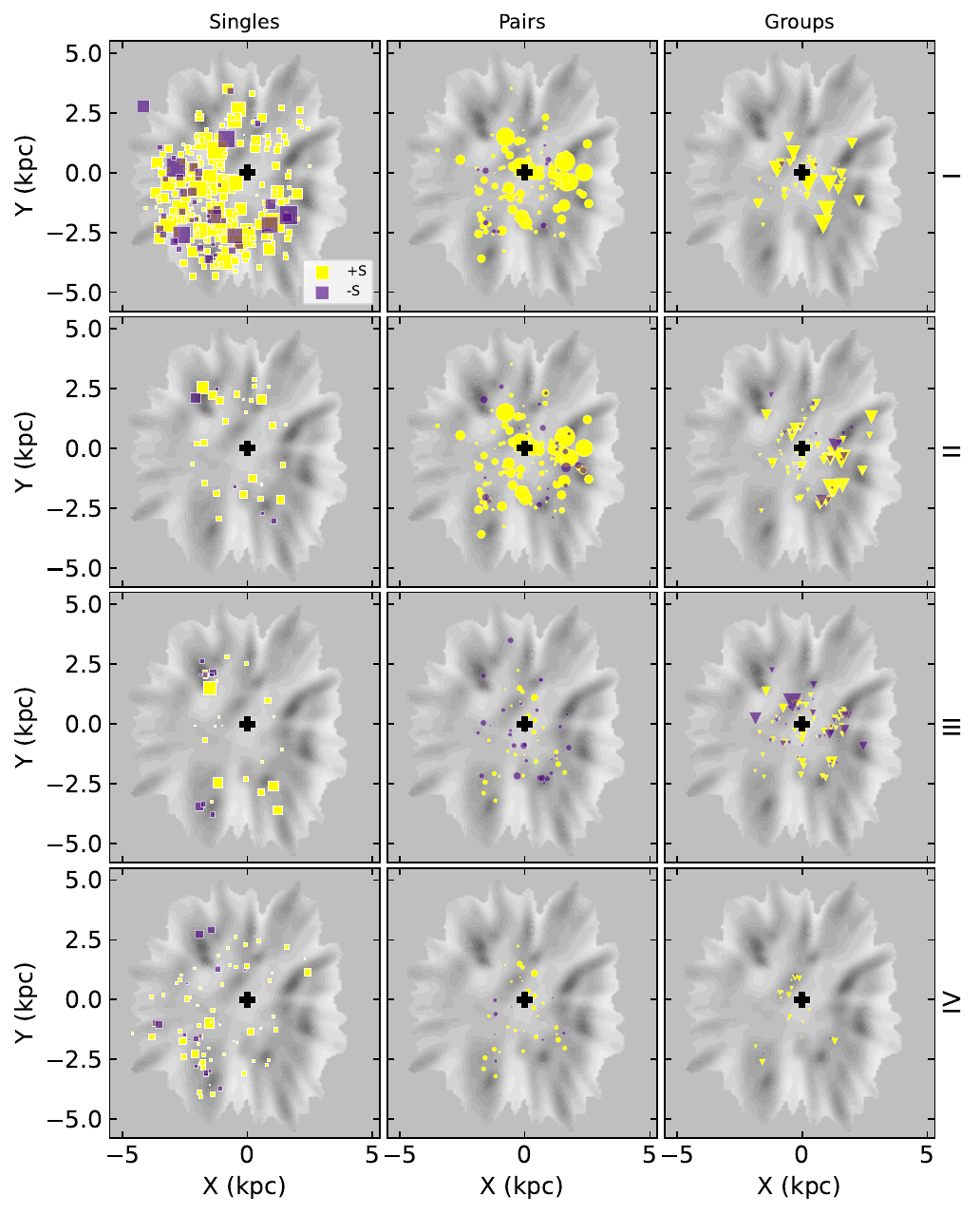}
\caption{
Face-on view of the stellar overdensity map, analogous to Fig. \ref{fig:overdensity}. The top panels show the first quadrant of Fig. \ref{fig:pairs}, followed by the second, third, and fourth quadrants. Clusters with mass segregation (+S) are yellow, while clusters with inverse mass segregation (-S) are violet. The size of the marker corresponds to the total mass.
}
\label{fig:mapa2}
\end{figure}

\clearpage
\end{appendix}


\end{document}